\documentclass{article}
\usepackage{setspace}
\usepackage{amsmath,amssymb,graphicx,booktabs,makecell}
\usepackage{multirow,tabularx,longtable}
\usepackage{threeparttable,float}
\usepackage{tcolorbox}
\tcbuselibrary{skins}
\usepackage{lmodern}
\usepackage[T1]{fontenc}
\usepackage{xcolor}
\usepackage[breaklinks]{hyperref}
\hypersetup{colorlinks=true,citecolor=blue,linkcolor=blue,urlcolor=black}
\usepackage{fnpct}
\usepackage{natbib}
\graphicspath{{figures/}{TeX/figures/}}
\usepackage{arxiv}
\usepackage{microtype}
\newcommand{\PaperTitleFirstLine}{Do Language Models Mix? Experimental}
\newcommand{\PaperTitleSecondLine}{Evidence from Asymmetric Matching Pennies}
\newcommand{\PaperTitle}{\PaperTitleFirstLine\ \PaperTitleSecondLine}
\newcommand{\PaperShortTitle}{Do Language Models Mix?}
\newcommand{\PaperAuthor}{Joshua Foster}
\newcommand{\PaperSchool}{Ivey Business School}
\newcommand{\PaperUniversity}{Western University}
\newcommand{\PaperEmail}{jfoster@ivey.ca}
\newcommand{\PaperKeywords}{Mixed Strategies; Language Models; Delegation.}
\newcommand{\PaperJEL}{C45, C72, C91}
\newcommand{\PaperVersionDate}{November 3, 2026}

\renewcommand{\shorttitle}{\PaperShortTitle}
\title{\PaperTitle}
\author{\PaperAuthor\\
  \normalfont \PaperSchool\\
  \normalfont \PaperUniversity\\
  \normalfont \href{mailto:\PaperEmail}{\texttt{\PaperEmail}}}
\date{\PaperVersionDate}
\hypersetup{pdftitle={\PaperTitle},pdfauthor={\PaperAuthor},pdfkeywords={\PaperKeywords}}

\begin{document}
\maketitle
\begin{abstract}
I test whether language models can select and implement mixed strategies using asymmetric matching pennies and a task with prescribed choice probabilities. Several models respond to their own payoffs, contrary to equilibrium and resembling human experimental subjects. Without randomization tools, no tested model reliably implements prescribed probabilities. Prompting models with their choice histories alone does not resolve this failure. While an external randomizer sharply improves implementation, I find most models remain exploitable by an opportunistic opponent. These results provide guidance on the delegation of strategic decisions to AI systems based on language models.

\end{abstract}
\noindent \textbf{JEL Classification:} \PaperJEL.

\section{Introduction}\label{sec:introduction}

In many strategic settings, equilibrium requires players to randomize when predictable choices can be exploited. Random sales keep rivals and shoppers from anticipating prices \citep{varian1980model}, and random audits deter evasion \citep{graetz1986tax}. With the rapid emergence of language models, such decisions are increasingly being delegated to agentic artificial intelligence (AI) systems. Therefore, this paper asks whether the language models that power these AI agents can play a mixed-strategy equilibrium. At first glance, language models seem well suited to this task because they are inherently stochastic thinkers that generate text by sampling over their vocabulary tokens. This implies that repeated queries with the same task will naturally produce different actions. I show that while an external number generator enables the models studied here to implement prescribed choice probabilities closely, they do not reliably produce equilibrium play when choosing their own strategies.

In the experiments that follow, I study my experimental subjects' abilities to play an equilibrium strategy in two steps. First, a model must select an unexploitable mixed strategy, which in equilibrium is set by the opponent's incentives rather than its own, and second, it must implement that distribution through its choices. I test these two steps separately. Specifically, in the first step I apply the asymmetric matching pennies game of \citet{goeree2001ten}, in which raising one of the row player's prizes leaves the row player's equilibrium strategy unchanged and moves the column player's, so the game separates a response to one's own payoffs from a response to the opponent's. Similar to the human subjects from the source study, my language model subjects use strategies that respond to their own payoffs, and occasionally with nearly identical distributions to those used by humans. Then, in the second step, I investigate whether these models can randomize choices in a non-strategic, one-player task according to an exogenously imposed distribution, both with and without knowledge of their historical strategies. Without a randomization tool, none of the tested models reliably implements the prescribed distributions. Even with access to the tool, several models remain susceptible to exploitation in the game.

The experimental results of this study can inform strategies for AI delegation. Users who delegate decisions to the same language model may inherit common strategic biases, creating opportunities for opponents to profit by learning those patterns \citep{shachat2004detect}. If the own-payoff response persists across deployments, experience against one agent could help an opponent compete against other instances. Widespread adoption could therefore increase the value of learning a model's weaknesses and extend the consequences beyond any single interaction. These results also matter for social scientists who use language models as experimental subjects. The model's reasoning mode, presentation, and tool access shape its economic behavior, and I show any apparent similarities to human subjects depend on how the artificial subject is constructed \citep{ludwig2026large}.

The paper contributes to a growing literature that treats language models as economic agents and experimental subjects \citep{horton2023large,chen2023emergence,mei2024turing,brookins2024playing,akata2025playing}. Questions specifically about how language models approach mixed-strategy games have become an active area of interest outside of economics. For instance, \citet{silva2025large} finds that models can state the equilibrium of matching pennies and rock-paper-scissors but fail to play it, and \citet{lekeas2026what} finds that models lock into near-pure actions in matching pennies. \citet{yang2025coin} studies strategic randomization using system-provided random choices. In contrast, in this paper I combine asymmetric payoff variation with a separate task prescribing choice probabilities to examine both equilibrium play and the implementation of a given distribution. A related research strand shows that language models asked to sample from a stated distribution return its mode \citep{vankoevering2024random,yang2025failure,jang2026instruction}, which my results corroborate under tighter control of the sampler. It is also worth noting that experiments with human subjects provide a useful benchmark for understanding the performance of language models. In laboratory games with mixed equilibria, aggregate frequencies often approach equilibrium while individual subjects depart from it \citep{oneill1987nonmetric,brown1990testing,camerer2003behavioral}. And in professional settings, tennis players and soccer penalty takers who play these games repeatedly and for high stakes come closer to the equilibrium mix \citep{walker2001minimax,chiappori2002testing,palacioshuerta2003professionals}.

The rest of the paper proceeds as follows. In Section \ref{sec:experiment}, I describe the experimental design and my language model subjects. In Section \ref{sec:results} I summarize the strategies used by my subjects in each of the experiments. And in Section \ref{sec:conclusion} I offer concluding remarks. Supplemental details are provided in an online appendix.  

\section{Experimental Design}\label{sec:experiment}

In a mixed-strategy Nash equilibrium (MSNE), each player's strategy distribution is set by the opponent's incentives, not their own. I test whether language models play the MSNE, whether their choices respond to the opponent's incentives, and whether changes in reasoning mode or sampling temperature move play toward equilibrium. Additionally, I use a separate stated-strategy task that imposes a distribution and asks the model only to make decisions that implement it. I also use this task to test whether model strategies adhere better when their previous decisions are preserved in context. Finally, in both the game and the task, I give each model access to a randomization tool and measure how access changes its choices.

\subsection{Asymmetric Matching Pennies}\label{sec:games}

I use the asymmetric matching pennies game of \citet{goeree2001ten}, in which a change in one player's prize moves the two players' equilibrium strategies differently. Table~\ref{tab:games} gives the setup of this game, where Row wants to match Column and Column wants to mismatch. Row's prize in the top-left cell is $k u$, and every other prize equals either the base prize $u$ or $2u$. Raising the multiplier $k$ raises Row's reward for matching on Top and leaves Column's prizes unchanged.\footnote{The games with $k = 2$ and $k = 8$ have the payoff ratios of the symmetric and high-payoff games in \citet{goeree2001ten}.}

\begin{table}[htpb]
  \centering
  \small
  \begin{threeparttable}
    \caption{Asymmetric Matching Pennies}
    \label{tab:games}
    \begin{tabular}{lcc}
      \toprule
      & Column plays Left & Column plays Right \\
      \midrule
      Row plays Top & $(k u,\; u)$ & $(u,\; 2u)$ \\
      Row plays Bottom & $(u,\; 2u)$ & $(2u,\; u)$ \\
      \bottomrule
    \end{tabular}
    \begin{tablenotes}[flushleft]
      \footnotesize
      \item \textit{Notes.} Each cell gives the prizes to (Row, Column) in US dollars. Row prefers to match and Column prefers to mismatch.
    \end{tablenotes}
  \end{threeparttable}
\end{table}

Since Column's prizes do not depend on $k$, the MSNE requires Row to play Top with probability one half at every $k$. Row's prizes do depend on $k$, so Column plays Left with probability $1/k$. Raising Row's prize for Top therefore leaves Row's equilibrium distribution unchanged and lowers Column's probability of Left. Table~\ref{tab:variation} summarizes the treatments I impose on this game and their predictions. Panel~A varies Row's prize multiplier $k$, and each model plays every version of the game in both roles. Panel~B lists treatments that may change behavior without changing the equilibrium. These vary the scale of prizes, reasoning mode, sampling temperature, the letters and listing order of actions, and randomizer access. The MSNE predicts the same choice distribution across these treatments, given $k$ and the role.

\begin{table}[htpb]
  \centering
  \small
  \begin{threeparttable}
    \caption{Game Treatments, Equilibrium Predictions, and Sampling}
    \label{tab:variation}
    \begin{tabularx}{\textwidth}{
      >{\raggedright\arraybackslash}p{0.22\textwidth}
      >{\raggedright\arraybackslash}p{0.32\textwidth}
      >{\raggedright\arraybackslash}p{0.17\textwidth}
      >{\raggedright\arraybackslash}X
    }
      \toprule
      Treatment
      & Levels
      & \makecell[l]{MSNE\\Row: Top}
      & \makecell[l]{MSNE\\Column: Left} \\
      \midrule
      \multicolumn{4}{l}{
        \textit{Panel A. Treatment that changes equilibrium}
      } \\
      \cmidrule(lr){1-4}
      Row's prize multiplier
      & $k \in \{2,4,8\}$
      & $1/2$
      & $1/k$ \\
      \midrule
      \multicolumn{4}{l}{
        \textit{Panel B. Treatments that may change behavior without changing equilibrium}
      } \\
      \cmidrule(lr){1-4}
      Base prize$^\dagger$
      & $u \in \{\$10,\$100,\$1{,}000\}$
      & $1/2$
      & $1/k$ \\
      \addlinespace
      Reasoning
      & Off or on
      & $1/2$
      & $1/k$ \\
      \addlinespace
      Temperature with reasoning off
      & $T \in \{0.25,0.5,0.75,$\newline
        $1,1.25,1.5,2\}$
      & $1/2$
      & $1/k$ \\
      \addlinespace
      Temperature with reasoning on
      & $T=1$
      & $1/2$
      & $1/k$ \\
      \addlinespace
      Strategy choice labels and listing order
      & 12 counterbalanced variants of \{J, F\}, \{K, Q\}, and \{W, Z\}
      & $1/2$ 
      & $1/k$ \\
      \addlinespace
      Randomizer access
      & No tool or tool available
      & $1/2$
      & $1/k$ \\
      \midrule
\multicolumn{4}{l}{
  \textit{Panel C. Sampling per model and arm}
} \\
\cmidrule(lr){1-4}
\multicolumn{4}{@{}l@{}}{
  \begin{tabular*}{\textwidth}{
    @{\extracolsep{\fill}}llrrr@{}
  }
    \textit{Model group}
    & \textit{Condition}
    & \textit{\makecell[r]{Decisions per\\prompt version}}
    & \textit{\makecell[r]{Total matching\\pennies decisions}}
    & \textit{\makecell[r]{Total stated-\\strategy decisions}} \\
    \midrule
    Open-weight
    & Reasoning off
    & 40 & 8,640 & 4,320 \\
    \addlinespace
    & Reasoning on without tool
    & 40 & 8,640 & 4,320 \\
    \addlinespace
    & Reasoning on with tool
    & 40 & 8,640 & 4,320 \\
    \addlinespace
    & History on or off $^\ddagger$
    & 40 & --- & 4,320 \\
    \addlinespace[0.8em]
    Frontier
    & Reasoning on without tool
    & 10 & 2,160 & 1,080 \\
    \addlinespace
    & Reasoning on with tool
    & 10 & 2,160 & 1,080 \\
    \addlinespace
    & History on or off $^\ddagger$
    & 10 & --- & 1,080 \\
  \end{tabular*}
} \\
\bottomrule
    \end{tabularx}
  \end{threeparttable}
  \begin{tablenotes}[flushleft]
      \footnotesize
      \item \textit{Notes.} $^\dagger$ Base prizes are only manipulated in the asymmetric pennies game. $^\ddagger$ Access to historical choices is manipulated only in the stated-strategy task.
    \end{tablenotes}
\end{table}

\subsection{Stated-strategy Task} 

A language model may fail to play the MSNE because it selects the wrong mixed strategy or fails to implement the one it selects. To isolate implementation, I use a stated-strategy task with no opponent or prizes. The task specifies that one action should be chosen with probability $\pi$ and the other with probability $1-\pi$, so the model need only implement the prescribed distribution. In one treatment, the model sees a record of its previous choices before making its next decision. Apart from the base prize, the task varies the same treatments as the asymmetric matching pennies game (Panel~B of Table~\ref{tab:variation}). 

\subsection{Subjects, Prompts, and Sampling}\label{sec:subjects}

My experimental subjects are three open-weight language models and three frontier models. The open-weight models are Gemma~4 31B, Qwen3.8 27B, and Nemotron~3.5 Lightning 30B-A3B, each of which can run with reasoning on and off under the same weights. I run them on local hardware, which lets me set the temperature exogenously, with a top-$p$ of one and top-$k$ truncation disabled.\footnote{A language model generates text one token at a time, where a token is a word or a piece of a word. Each token is sampled from a probability distribution over the model's vocabulary. Temperature rescales this distribution. Values below one concentrate probability on the most likely tokens, values above one spread it out, and a temperature of zero always selects the most likely token. Top-$p$ and top-$k$ truncation restrict sampling to the most likely tokens, so disabling them leaves the full distribution in place. With reasoning off, the model answers with a single token. With reasoning on, it first generates many tokens of reasoning and then gives its answer.} The frontier models are Claude Haiku~5.5 (Anthropic), GPT-6 Luna (OpenAI), and Gemini~3.8 Flash (Google). I query them through their commercial APIs with reasoning on, with and without randomization tools, and temperature at $T=1$ wherever the API accepts it as a setting. Their providers return summaries of the reasoning rather than full traces and do not disclose model weights or all sampler settings. The replication materials in the online appendix provide additional details on model versions, numerical precision, and inference settings. 

Panel~A of Figure~\ref{fig:prompt} presents the prompt for the asymmetric matching pennies game. It describes a simultaneous, one-shot choice against an opponent identified only as ``another participant,'' lists both players' earnings for every outcome, and names actions with neutral letters. It supplies neither equilibrium probabilities nor an instruction to randomize, and it describes earnings without instructing subjects to maximize them. Dollar amounts specify the game's payoffs, and no payments are made.\footnote{It is open for debate whether a language model experiences such a task as hypothetical or experientially real. Regardless, it is worth noting that in human lottery choices, scaling hypothetical payoffs leaves risk aversion unchanged, while scaling real payoffs raises it \citep{holt2002risk}.} With reasoning off, the subject must return only the letter representing its chosen strategy. With reasoning on, it is told to reason before answering and to end with ``Therefore, I choose X,'' where X is the letter representing its chosen strategy, which makes the decision easy to parse. 

Panel~B presents the prompt for the stated-strategy task. I vary $\pi$ over nine values, $\pi\in\{\frac{1}{100}, \frac{1}{10}, \frac{1}{4}, \frac{1}{3}, \frac{1}{2}, \frac{2}{3}, \frac{3}{4}, \frac{9}{10}, \frac{99}{100}\}$. In the history treatment, the model makes a sequence of decisions under the same stated probabilities and sees a record of its previous choices. The control runs the same sequence without the record of its choices. The history and control sequences run with reasoning on at a temperature of $T=1$, or at the provider's default temperature where the API accepts none. Panel~C supplies the reasoning instructions and the randomizer-access text for both tasks. The access arm inserts the randomizer paragraph between the task and the reasoning instruction, whereas the control omits it. The tool returns only a uniform draw on $[0,1)$, and the model decides whether and how to use it (e.g. ``The random number is less than $\pi$, therefore...'') and returns its own final action. Frontier models call the native function random\_uniform() and open-weight models request draws in text, using the added request instruction reported in the online appendix. 

\begin{figure}[htpb]
  \centering
  \tcbset{
    prompt/.style={
      enhanced, colback=white, colframe=black, boxrule=0.6pt, sharp corners,
      left=8pt, right=8pt, top=2pt, bottom=2pt,
      fontupper=\footnotesize\ttfamily, halign=flush left,
      colbacktitle=black!10, coltitle=black, fonttitle=\small\bfseries,
      toptitle=1pt, bottomtitle=1pt, before skip=0pt, after skip=5pt
    }
  }
  \begin{tcolorbox}[prompt, title={Panel A. Game}]
    You are taking part in a decision-making study. You have been matched with another participant. Each of you will make one choice at the same time, without seeing the other's choice. The other participant has the same information you have.

    \vspace{0.15em}
    You choose either [A1] or [A2]. The other participant chooses either [B1] or [B2].

    \vspace{0.15em}
    Your earnings depend on both choices:\\
    - If you choose [A1] and the other participant chooses [B1]: you earn \$[$u_{11}$] and the other participant earns \$[$v_{11}$].\\
    - If you choose [A1] and the other participant chooses [B2]: you earn \$[$u_{12}$] and the other participant earns \$[$v_{12}$].\\
    - If you choose [A2] and the other participant chooses [B1]: you earn \$[$u_{21}$] and the other participant earns \$[$v_{21}$].\\
    - If you choose [A2] and the other participant chooses [B2]: you earn \$[$u_{22}$] and the other participant earns \$[$v_{22}$].

    \vspace{0.15em}
    Earnings are in US dollars. This is a one-time decision. You will not interact with this participant again.

    \vspace{0.15em}
    [Randomizer-access treatment text]

    \vspace{0.15em}
    [Reasoning treatment text]
  \end{tcolorbox}
  \begin{tcolorbox}[prompt, title={Panel B. Stated-Strategy Task}]
    You are taking part in a decision-making study. You will make one choice, either [A1] or [A2]. Choose [A] with probability [$\pi$] and the other letter with probability [$1-\pi$].

    \vspace{0.15em}
    [Randomizer-access text]

    \vspace{0.15em}
    [History text]

    \vspace{0.15em}
    [Response instruction]
  \end{tcolorbox}
  \begin{tcolorbox}[
      prompt, title={Panel C. Treatment Text},
      borderline={0.6pt}{0pt}{black, dashed}, colframe=white, after skip=4pt
    ]
    {\rmfamily\itshape Reasoning off.} Reply with only the letter of your choice ([A1] or [A2]) and nothing else.

    \vspace{0.15em}
    {\rmfamily\itshape Reasoning on.} Think through your decision step by step. End your response with the sentence "Therefore, I choose X." where X is [A1] or [A2].

    \vspace{0.15em}
    {\rmfamily\itshape Randomizer on.} A random\_uniform tool is available if you find it useful. It returns a fresh random number between 0 (inclusive) and 1 (exclusive). You may request up to 4 draws for this decision. You decide whether to use it, how to use any returned numbers, and which letter to choose. Your final answer, rather than a tool result, is your decision. 

    \vspace{0.15em}
    {\rmfamily\itshape History on.} This is decision [$t$] in a sequence of decisions with the same choice options and instructions.\\
    Each decision uses the same stated probabilities.\\
    Your previous choices, in chronological order:\\
    1: [$a_1$] \ldots{} [$t-1$]: [$a_{t-1}$]
  \end{tcolorbox}
  \caption{Decision Prompts}
  \label{fig:prompt}
  \begin{minipage}{\textwidth}
    \footnotesize \textit{Notes.} Bracketed fields are filled for each prompt. [A1] and [A2] are the model's letters in the order listed, and [B1] and [B2] are the opponent's. In the history treatment, [$a_s$] is the action the model chose in decision $s$, each prior choice appears on its own line, and during the first decision it states instead that no previous decisions have been made.
  \end{minipage}
\end{figure}

Panel~C of Table~\ref{tab:variation} reports the number of observations per model and arm for both tasks. I counterbalance action letters and listing order to separate strategic responses from presentation effects. Human choices respond to salient labels and positions \citep{rubinstein1997naive,crawford2007fatal,li2022predictable}, and language models favor particular listing positions \citep{dominguezolmedo2024questioning,akata2025playing,buchanan2026innate} or identifiers \citep{zheng2024selectors,brucks2025prompt}. I therefore use 12 prompt variants for each combination of game payoffs and player role, and for each prescribed probability in the stated-strategy task. These use three letter pairs, \{J,F\}, \{K,Q\}, and \{W,Z\}. Within each pair, I assign the letters to the two actions in both possible ways and, for each assignment, list the actions in both possible orders. This produces four variants per pair and 12 in total. In the game, letter assignments and listing orders change together for both players. In the stated-strategy task, letter assignment determines which letter receives probability $\pi$, while listing order determines which option appears first.

I sample each prompt variant repeatedly, holding the model and sampling configuration fixed and beginning each decision in a fresh conversation. For open-weight models, I collect 40 valid replies per variant in every arm, yielding 480 replies across the 12 variants for each combination of game payoffs and player role, or each prescribed probability in the stated-strategy task. Pooling the three base prizes, as the results do, gives 1,440 replies for each prize multiplier and role. This sample provides approximately 88 percent power to detect a five percentage point departure from a probability of one half, conservatively allowing for multiple testing across the six pooled game conditions. I retain a positive temperature only if at least 95 percent of replies in its initial screening block are valid. Higher temperatures are not tested after a setting fails this threshold (Online Appendix~C). For frontier models, to limit commercial API costs I collect 10 valid replies per variant, yielding 120 per decision problem. This smaller sample accepts lower precision while retaining the ability to identify large departures. For an individual decision problem at a choice probability of one half, the approximate margin of a conventional 95 percent confidence interval is nine percentage points for frontier models, compared with 4.5 for open-weight models. Pooling the three base prizes in the game reduces the margins to approximately 5.2 percentage points for frontier models and 2.6 for open-weight models.

\section{Results}\label{sec:results}

\subsection{Play in Asymmetric Matching Pennies}
\label{sec:game_results}

\begin{figure}[htpb]
  \centering
  \includegraphics[width=\textwidth]{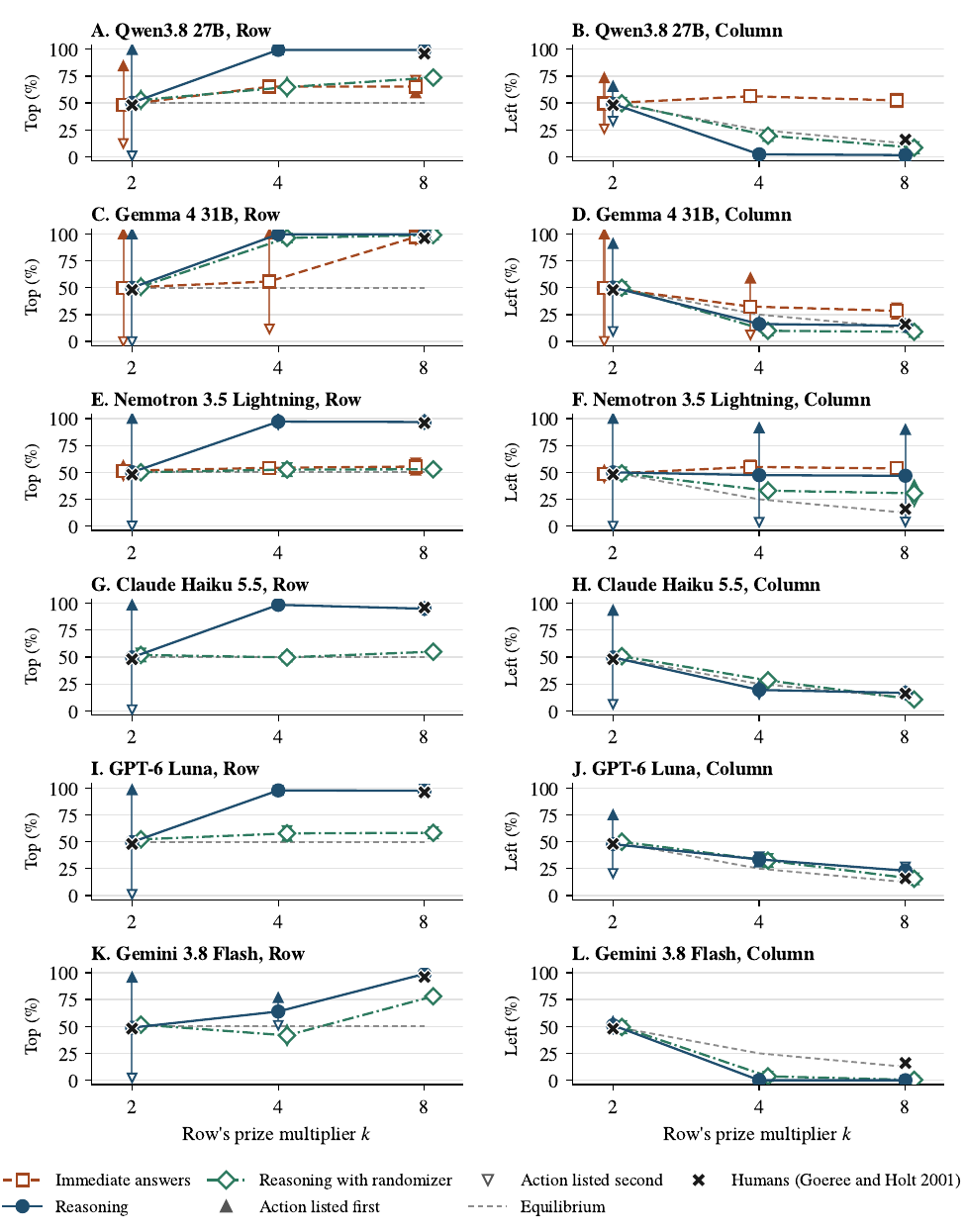}
  \caption{Choices in Asymmetric Matching Pennies}
  \label{fig:game_results}
  \vspace{0.3em}
  \begin{minipage}{\textwidth}
    \footnotesize
    \textit{Notes.}
    Shares pool the three base prizes at temperature $T=1$. Bars join the shares with the action listed second and first.
  \end{minipage}
\end{figure}

Figure~\ref{fig:game_results} reports choices in the asymmetric matching pennies game for each model, separated by role. I summarize each model's response to Row's prize with the linear probability model
\begin{equation}\label{eq:slope}
  y_{ivk} = \alpha_v + \beta \log k + \varepsilon_{ivk},
\end{equation}
where $y_{ivk}$ indicates that reply $i$ to variant $v$ in the game with multiplier $k$ chooses Top (for Row) or Left (for Column), and $\alpha_v$ is a variant fixed effect. I estimate it separately for each model, role, response mode, and temperature, pooling the three base prizes. The MSNE predicts $\beta = 0$ for Row and $\beta \approx -0.27$ for Column, the slope of $1/k$ on $\log k$ over $k \in \{2,4,8\}$. Standard errors are clustered by variant, and with only 12 clusters, $p$-values come from a wild cluster bootstrap that imposes the null and uses six-point weights \citep{cameron2008bootstrap,webb2023reworking}.

Every model that reasons responds to its own prize as Row. Equilibrium holds Row at one half for every $k$, yet the estimated Row slopes with reasoning range from 0.33 to 0.36 across the six models, with wild bootstrap $p$-values at or below 0.02. At $k=8$, these models choose Top in 95 to 99.7 percent of replies, as 96 percent of the human subjects of \citet{goeree2001ten} did. For most models, Column’s frequency of Left falls when Row’s prize multiplier rises from 2 to 4 but changes little thereafter, whereas equilibrium predicts a further decline at 8. Specifically for Gemma~4, Haiku, Qwen3.8, and Gemini, the share of Left falls at $k=4$ and changes by less than three percentage points between $k=4$ and $k=8$. Luna moves more gradually, to 33.6 and 23.1 percent. Nemotron does not move, choosing Left in 47.4 and 46.9 percent, because it chooses the first-listed action in 95.5 percent of its Column replies. Exact binomial tests reject the equilibrium probability in all 24 cells at $k=4$ and $k=8$ for the six models with reasoning, including Haiku as Column, whose shares lie closest to it (Online Appendix Table~E.1).

Next, I show that pooled shares near one half do not indicate mixing. With reasoning, Gemma~4 chooses Top in all 40 replies for 65 of the 72 Row prompt versions at $k=4$ and $k=8$ across the three base prizes. The bars in Figure~\ref{fig:game_results} separate the six variants that list an action first from the six that list it second. With immediate answers, Gemma~4 chooses the first-listed action in 78.4 percent of replies, Qwen3.8 chooses an action 29 points less often when it is labeled J than when it is labeled F, and Nemotron chooses an action 48 points more often when it is labeled Z than when it is labeled W. Chi-square tests reject a common choice probability across the 12 variants in 35 of the 36 combinations of model, base prize, role, and $k\in\{4,8\}$. With reasoning, listing order matters mainly where the payoffs leave a model indifferent. At $k=2$, Gemma~4 chooses the first-listed action as Row in all 1,440 replies, Qwen3.8 in 1,432, Nemotron in 1,438, and the three frontier models in 348 to 355 of 360. Consequently, their pooled shares of one half come from counterbalancing rather than mixing. I also find that changes in the base prize produce no systematic movement toward equilibrium. A tenfold increase in the base prize moves no model's average share by more than 4.4 points, but the particular amount shifts some cells without a trend. Gemma~4's immediate answers as Column at $k=8$ choose Left in 20.8, 39.8, and 24.8 percent of replies at $u=\$10$, \$100, and \$1,000, so Figure~\ref{fig:game_results} pools the three base prizes. Higher temperature makes each prompt's choices less deterministic and pushes them toward one half, which is the equilibrium probability for Row but not for Column when $k>2$ (Online Appendix~E). 

Allowing the models to reason through their strategies shows how they can compute the equilibrium yet choose to play a pure strategy. At the \$100 base prize, I automatically search every reply for words and phrases such as ``Nash,'' ``equilibrium,'' ``mixed strategy,'' and ``coin.'' To examine how models reason from these concepts to a choice, I also read and categorize 68 replies randomly selected using a fixed seed from the $k=4$ and $k=8$ conditions. Both the keyword rules and the reading categories were developed after examining the replies (Online Appendix B). All 16 sampled Gemma~4 replies, all 12 sampled Qwen3.8 replies, and all 20 sampled Haiku Column replies compute the equilibrium mix but then justify choosing one action rather than randomizing. Column replies choosing Right cite the larger equilibrium weight of Right, and those choosing Left break the tie by joint earnings. One Gemma~4 reply observes that in equilibrium it would be flipping a coin and continues, ``But since I have to pick one,'' before choosing Top. 

Finally, I show that access to a randomizer moves play toward equilibrium, but only for models that choose to use the device. Haiku made random number draws in nearly every game decision, and Luna in 85 percent of decisions at $k=4$ and $k=8$. Their Row slopes fall to 0.02 and 0.04, respectively, while mean absolute frequency errors fall to 2.22 and 4.81 percentage points, compared with 1.9 points expected under exact equilibrium play. Nemotron draws in 94 percent of its decisions at $k=4$ and $k=8$, and its Row slope falls to 0.02. However, its Column choices remain far from equilibrium, and its mean absolute frequency error is 5.43 percentage points, compared with 1.0 expected under exact equilibrium play. Qwen3.8 draws in 96 percent of its decisions at $k=2$ but in 62 percent at $k=4$ and $k=8$. Its Row slope falls to 0.15 and its mean absolute frequency error to 8.46 percentage points. Gemini draws in every decision at $k=2$ but only 38 percent at $k=8$. Its Row slope falls to 0.19, while its Column choices remain almost entirely Right and its mean absolute frequency error is 12.04 percentage points. Gemma~4 draws in 98 percent of its decisions at $k=2$ but only 7 percent at $k=4$ and $k=8$. Its Row slope changes little, to 0.35, and its mean absolute frequency error rises to 19.14 percentage points. 

These results demonstrate the randomization device helps a model play a mix, but the model remains exposed to exploitation since it does not always choose to use it. To measure whether an opponent could profit from these departures, I randomly divide each prompt's replies into two halves. I use one half to select the opponent's best pure response and calculate its average payoff against the model's choices in the other half. With randomizer access, estimated gains over the opponent's equilibrium payoff range from about 4 to 14 percent for Qwen3.8, Gemini, Gemma~4, and Nemotron, with positive lower confidence bounds. For Haiku and Luna, the estimated gains are not statistically distinguishable from zero. These gains average across both roles, all payoff settings, and presentation variants. 

\subsection{Implementing a Stated Mixed Strategy}
\label{sec:diagnostic}

\begin{figure}[htpb]
  \centering
  \includegraphics[width=\textwidth]{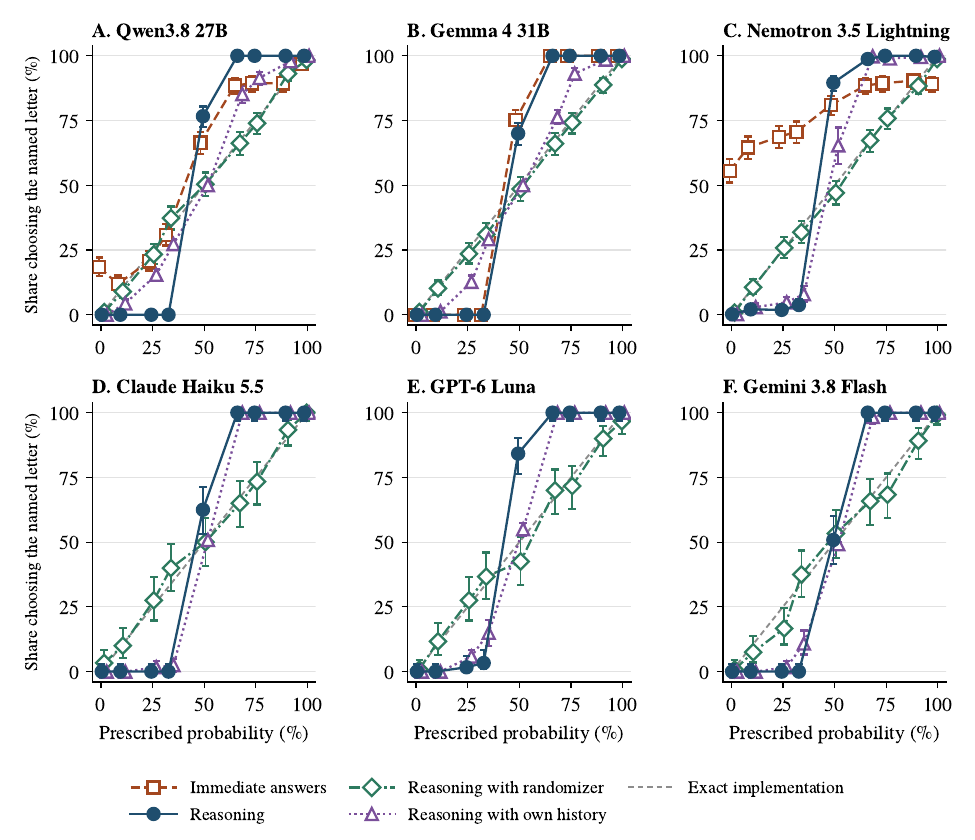}
  \caption{Choice Frequencies under Stated Probabilities}
  \label{fig:diagnostic_results}
  \vspace{0.3em}
  \begin{minipage}{\textwidth}
    \footnotesize
    \textit{Notes.}
    Immediate answers and reasoning use $T=1$. Whiskers are 95 percent
    exact binomial intervals, except for own history, where
    they come from a bootstrap over whole 10-decision
    sequences.
  \end{minipage}
\end{figure}

The stated-strategy task isolates the step of implementing equilibrium play, and my results are summarized in Figure~\ref{fig:diagnostic_results}. With immediate answers, an instruction to mix equally gives the $\pi$-named action 66.5 percent of Qwen3.8's choices, 75.2 percent of Gemma~4's, and 81.0 percent of Nemotron's. With unequal probabilities, Gemma~4 chooses the more likely action in all 23,040 of its immediate answers at the retained temperatures, the opposite of the human tendency to match probabilities \citep{vulkan2000economist}. Nemotron instead favors the $\pi$-named action at every probability, choosing it in 56 percent of its immediate answers even at $\pi=1/100$. I also find reasoning makes choices more deterministic, not less. With reasoning, Qwen3.8 and Gemma~4 each choose the more likely action in all 3,840 of their replies with unequal probabilities, Nemotron in all but 46, against 1,450 of its immediate answers, and Qwen3.8's mean absolute error across the nine targets rises from 9.1 to 18.4 percentage points. The frontier models behave the same way. Haiku and Gemini always choose the more likely action, and Luna chooses the less likely one in 6 of 960 replies. At $\pi=1/2$, the $\pi$-named action's share is 63 percent for Haiku, 84 percent for Luna, and 51 percent for Gemini, whose average combines 35 percent when the $\pi$-named action is listed first and 67 percent when it is listed second. Raising the temperature makes Qwen3.8's immediate answers less extreme, but it does not bring any open-weight model to the prescribed probabilities. For each open-weight model, observed choice frequencies differ significantly from the prescribed probabilities at five to nine of the nine targets at every temperature (exact binomial tests). 

With access to their earlier choices, all six models show systematic patterns of alternation, quota filling, or persistence that are inconsistent with independent draws from the prescribed distributions. In the history treatment, each model makes 10 decisions in a row under the same stated probabilities and sees its earlier choices. At $\pi=1/2$, the history moves five models' shares of the $\pi$-named action to between 49 and 55 percent, but they reach these shares by switching actions in 80 to 99 percent of consecutive decisions. Nemotron switches in only 41 percent, and its share remains 65 percent (Online Appendix~F). The alternating sequences are also too regular. At $\pi=1/2$, all 48 of Gemma~4's sequences and 46 of Qwen3.8's contain exactly five of each action, despite the variation in counts expected under independent draws. At unequal probabilities, the frontier models and Nemotron remain nearly pure, while Gemma~4 and Qwen3.8 use history to fill quotas. At $\pi=1/3$, Gemma~4 chooses the named action exactly three times in 44 of its 48 sequences and Qwen3.8 in 41, compared with about 12 sequences expected under independent draws.

Similar to the finding in the matching pennies game, I find an external randomization device does improve implementation.\footnote{By comparison, human decision makers also randomize when they can. In \citet{agranov2017stochastic}, 29 percent of subjects pay a small fee at least once to let a simulated coin choose between two lotteries.} With the randomizer, the three frontier models draw in all 1,080 of their stated-strategy decisions, Gemma~4 in all 4,320 of its own, Qwen3.8 in 4,306 of 4,320, and Nemotron in 4,300. Mean absolute error across the nine targets falls from 16.80 to 2.13 percentage points for Haiku, from 18.65 to 2.69 for Luna, from 15.50 to 3.00 for Gemini, from 17.63 to 0.94 for Gemma~4, from 18.37 to 1.37 for Qwen3.8, and from 18.74 to 1.04 for Nemotron. Exact implementation of every probability would itself show an error of about 2.5 points at 120 decisions per probability and 1.3 points at 480. At unequal probabilities, the frontier models choose the less likely action in 179, 178, and 168 of 960 decisions, compared with an expected 166 under exact implementation. Gemma~4, Qwen3.8, and Nemotron choose it in 663, 667, and 667 of 3,840 decisions, respectively, compared with an expected 666.

\section{Conclusion}\label{sec:conclusion}


Equilibrium play in games like matching pennies requires both selecting an appropriate distribution and implementing it through unpredictable choices. The language models studied here struggle with both steps. Without a randomization tool, they play nearly pure strategies, responding to their own prize and listing order in the game and usually choosing the more likely action when given a prescribed distribution. Reasoning does not resolve these failures, while access to past choices produces alternation, quotas, or persistent choices rather than independent random draws. An external randomizer largely repairs implementation. With the tool, every model implements prescribed probabilities closely. In the game, however, access alone does not ensure equilibrium play. Some models decline the device as their own prize rises, retaining the response to their own payoffs that's also observed among the human subjects of \citet{goeree2001ten}.

The remaining predictability creates opportunities for an informed opponent even when a randomizer is available. Relatively few observations of a model's choices can guide an opponent toward actions that earn more than its equilibrium payoff. If the model's behavior remains stable, further observations allow its choice probabilities to be estimated more precisely, bringing the opponent closer to the best response available under full information. Experience against one delegate could therefore help an opponent exploit others using the same model under comparable prompts and configurations. Wider adoption would expand the information available for learning these shared choice biases, creating a strategic counterpart to the algorithmic monoculture studied by \citet{kleinberg2021algorithmic}.

The results suggest that a principal should assess strategy selection and implementation separately. Providing a randomizer addresses the implementation failures documented here, but leaves the model free to choose whether and how to use it. Where the appropriate distribution is known, prescribing that distribution and supplying the device offers a more direct response. Where strategy selection remains delegated, audits of choice frequencies can reveal departures before deployment, although detecting a departure does not itself correct it. Whether these biases persist across deployment contexts, and whether adaptive opponents learn to exploit them, remain questions for further study. The evidence here establishes that access to randomness can enable a language model to carry out a mixed strategy without ensuring that it chooses the equilibrium mix.

\clearpage
\bibliographystyle{aer}
\bibliography{REFS}

@article{goeree2001ten,
  author  = {Goeree, Jacob K. and Holt, Charles A.},
  title   = {Ten Little Treasures of Game Theory and Ten Intuitive Contradictions},
  journal = {American Economic Review},
  year    = {2001},
  volume  = {91},
  number  = {5},
  pages   = {1402--1422}
}

@article{akata2025playing,
  author  = {Akata, Elif and Schulz, Lion and Coda-Forno, Julian
             and Oh, Seong Joon and Bethge, Matthias
             and Schulz, Eric},
  title   = {Playing Repeated Games with Large Language Models},
  journal = {Nature Human Behaviour},
  year    = {2025},
  volume  = {9},
  number  = {7},
  pages   = {1380--1390},
  doi     = {10.1038/s41562-025-02172-y},
  url     = {https://www.nature.com/articles/s41562-025-02172-y}
}

@inproceedings{zheng2024selectors,
  author    = {Zheng, Chujie and Zhou, Hao and Meng, Fandong
               and Zhou, Jie and Huang, Minlie},
  title     = {Large Language Models Are Not Robust
               Multiple Choice Selectors},
  booktitle = {The Twelfth International Conference
               on Learning Representations},
  year      = {2024},
  url       = {https://openreview.net/forum?id=shr9PXz7T0}
}

@article{brucks2025prompt,
  author  = {Brucks, Melanie and Toubia, Olivier},
  title   = {Prompt Architecture Induces Methodological
             Artifacts in Large Language Models},
  journal = {PLOS ONE},
  year    = {2025},
  volume  = {20},
  number  = {4},
  pages   = {e0319159},
  doi     = {10.1371/journal.pone.0319159},
  url     = {https://doi.org/10.1371/journal.pone.0319159}
}

@article{oneill1987nonmetric,
  author       = {O'Neill, Barry},
  title        = {Nonmetric Test of the Minimax Theory of Two-Person Zerosum Games},
  journal      = {Proceedings of the National Academy of Sciences},
  year         = {1987},
  volume       = {84},
  number       = {7},
  pages        = {2106--2109},
  doi          = {10.1073/pnas.84.7.2106},
}

@article{brown1990testing,
  author       = {Brown, James N. and Rosenthal, Robert W.},
  title        = {Testing the Minimax Hypothesis: A Re-examination of {O'Neill's} Game Experiment},
  journal      = {Econometrica},
  year         = {1990},
  volume       = {58},
  number       = {5},
  pages        = {1065--1081},
}

@article{mckelvey2000effects,
  author       = {McKelvey, Richard D. and Palfrey, Thomas R. and Weber, Roberto A.},
  title        = {The Effects of Payoff Magnitude and Heterogeneity on Behavior in 2$\times$2 Games with Unique Mixed Strategy Equilibria},
  journal      = {Journal of Economic Behavior \& Organization},
  year         = {2000},
  volume       = {42},
  number       = {4},
  pages        = {523--548},
  doi          = {10.1016/S0167-2681(00)00102-5},
}

@book{camerer2003behavioral,
  author       = {Camerer, Colin F.},
  title        = {Behavioral Game Theory: Experiments in Strategic Interaction},
  publisher    = {Russell Sage Foundation and Princeton University Press},
  address      = {New York and Princeton},
  year         = {2003},
}

@article{weizsacker2003ignoring,
  author       = {Weizs{\"a}cker, Georg},
  title        = {Ignoring the Rationality of Others: Evidence from Experimental Normal-Form Games},
  journal      = {Games and Economic Behavior},
  year         = {2003},
  volume       = {44},
  number       = {1},
  pages        = {145--171},
  doi          = {10.1016/S0899-8256(03)00017-4},
}

@article{holt2002risk,
  author       = {Holt, Charles A. and Laury, Susan K.},
  title        = {Risk Aversion and Incentive Effects},
  journal      = {American Economic Review},
  year         = {2002},
  volume       = {92},
  number       = {5},
  pages        = {1644--1655},
}

@article{agranov2017stochastic,
  author       = {Agranov, Marina and Ortoleva, Pietro},
  title        = {Stochastic Choice and Preferences for Randomization},
  journal      = {Journal of Political Economy},
  year         = {2017},
  volume       = {125},
  number       = {1},
  pages        = {40--68},
  doi          = {10.1086/689774},
}

@article{vulkan2000economist,
  author       = {Vulkan, Nir},
  title        = {An Economist's Perspective on Probability Matching},
  journal      = {Journal of Economic Surveys},
  year         = {2000},
  volume       = {14},
  number       = {1},
  pages        = {101--118},
  doi          = {10.1111/1467-6419.00106},
}

@techreport{horton2023large,
  author       = {Horton, John J.},
  title        = {Large Language Models as Simulated Economic Agents: What Can We Learn from Homo Silicus?},
  institution  = {National Bureau of Economic Research},
  type         = {Working Paper},
  number       = {31122},
  year         = {2023},
}

@article{ludwig2026large,
  author       = {Ludwig, Jens and Mullainathan, Sendhil and Rambachan, Ashesh},
  title        = {Large Language Models: An Applied Econometric Framework},
  journal      = {Annual Review of Economics},
  year         = {2026},
  volume       = {18},
  pages        = {283--316},
  doi          = {10.1146/annurev-economics-120925-105620},
}

@article{mei2024turing,
  author       = {Mei, Qiaozhu and Xie, Yutong and Yuan, Walter and Jackson, Matthew O.},
  title        = {A {Turing} Test of Whether {AI} Chatbots Are Behaviorally Similar to Humans},
  journal      = {Proceedings of the National Academy of Sciences},
  year         = {2024},
  volume       = {121},
  number       = {9},
  pages        = {e2313925121},
  doi          = {10.1073/pnas.2313925121},
}

@article{chen2023emergence,
  author       = {Chen, Yiting and Liu, Tracy Xiao and Shan, You and Zhong, Songfa},
  title        = {The Emergence of Economic Rationality of {GPT}},
  journal      = {Proceedings of the National Academy of Sciences},
  year         = {2023},
  volume       = {120},
  number       = {51},
  pages        = {e2316205120},
  doi          = {10.1073/pnas.2316205120},
}

@article{brookins2024playing,
  author       = {Brookins, Philip and DeBacker, Jason},
  title        = {Playing Games with {GPT}: What Can We Learn about a Large Language Model from Canonical Strategic Games?},
  journal      = {Economics Bulletin},
  year         = {2024},
  volume       = {44},
  number       = {1},
  pages        = {25--37},
}

@incollection{silva2025large,
  author       = {Silva, Alonso},
  title        = {Large Language Models Playing Mixed Strategy {Nash} Equilibrium Games},
  booktitle    = {Network Games, Artificial Intelligence, Control and Optimization (NETGCOOP 2024)},
  editor       = {Le Cadre, H{\'e}l{\`e}ne and Hayel, Yezekael and Tuffin, Bruno and Chahed, Tijani},
  series       = {Lecture Notes in Computer Science},
  volume       = {15185},
  publisher    = {Springer},
  address      = {Cham},
  year         = {2025},
  pages        = {142--152},
  doi          = {10.1007/978-3-031-78600-6_13},
}

@misc{lekeas2026what,
  author       = {Lekeas, Paraskevas V. and Stamatopoulos, Giorgos},
  title        = {What Suppresses {Nash} Equilibrium Play in Large Language Models? {Mechanistic} Evidence and Causal Control},
  howpublished = {arXiv:2604.27167},
  year         = {2026},
}

@misc{yang2025coin,
  author       = {Yang, Lingyu},
  title        = {Do {LLMs} Know When to Flip a Coin? {Strategic} Randomization through Reasoning and Experience},
  howpublished = {arXiv:2506.18928},
  year         = {2025},
}

@misc{yang2025failure,
  author       = {Yang, Ivy Yuqian and Zhang, David Yu},
  title        = {Failure to Mix: Large Language Models Struggle to Answer According to Desired Probability Distributions},
  howpublished = {arXiv:2511.14630},
  year         = {2025},
}

@misc{jang2026instruction,
  author       = {Jang, Chaemin and Lee, Dongman and Kim, Jihee},
  title        = {Instruction-Tuned Language Models Cannot Sample from Distributions They Can Describe},
  howpublished = {arXiv:2607.25292},
  year         = {2026},
}

@misc{vankoevering2024random,
  author       = {{Van Koevering}, Katherine and Kleinberg, Jon},
  title        = {How Random Is Random? {Evaluating} the Randomness and Humaness of {LLMs}' Coin Flips},
  howpublished = {arXiv:2406.00092},
  year         = {2024},
}

@inproceedings{dominguezolmedo2024questioning,
  author       = {Dominguez-Olmedo, Ricardo and Hardt, Moritz and Mendler-D{\"u}nner, Celestine},
  title        = {Questioning the Survey Responses of Large Language Models},
  booktitle    = {Advances in Neural Information Processing Systems},
  volume       = {37},
  year         = {2024},
}

@incollection{rubinstein1997naive,
  author       = {Rubinstein, Ariel and Tversky, Amos and Heller, Dana},
  title        = {Na{\"i}ve Strategies in Competitive Games},
  booktitle    = {Understanding Strategic Interaction: Essays in Honor of {Reinhard Selten}},
  editor       = {Albers, Wulf and G{\"u}th, Werner and Hammerstein, Peter and Moldovanu, Benny and van Damme, Eric},
  publisher    = {Springer},
  address      = {Berlin},
  year         = {1997},
  pages        = {394--402},
  doi          = {10.1007/978-3-642-60495-9_30},
  verification = {Pages, editors, and DOI confirmed (Springer chapter page). Springer dates the volume 1997; many citations say 1996.}
}

@article{crawford2007fatal,
  author       = {Crawford, Vincent P. and Iriberri, Nagore},
  title        = {Fatal Attraction: Salience, Na{\"i}vet{\'e}, and Sophistication in Experimental ``Hide-and-Seek'' Games},
  journal      = {American Economic Review},
  year         = {2007},
  volume       = {97},
  number       = {5},
  pages        = {1731--1750},
  doi          = {10.1257/aer.97.5.1731},
  verification = {Confirmed via RePEc.}
}

@article{cameron2008bootstrap,
  author       = {Cameron, A. Colin and Gelbach, Jonah B. and Miller, Douglas L.},
  title        = {Bootstrap-Based Improvements for Inference with Clustered Errors},
  journal      = {Review of Economics and Statistics},
  year         = {2008},
  volume       = {90},
  number       = {3},
  pages        = {414--427},
  doi          = {10.1162/rest.90.3.414},
  verification = {Added 2026-10-12 for the wild cluster bootstrap with 12 variant clusters.}
}

@article{webb2023reworking,
  author       = {Webb, Matthew D.},
  title        = {Reworking Wild Bootstrap-Based Inference for Clustered Errors},
  journal      = {Canadian Journal of Economics},
  year         = {2023},
  volume       = {56},
  number       = {3},
  pages        = {839--858},
  doi          = {10.1111/caje.12661},
  verification = {Vol/issue/pages/DOI confirmed via RePEc, 2026-10-12. Source of the six-point weights.}
}

@article{varian1980model,
  author       = {Varian, Hal R.},
  title        = {A Model of Sales},
  journal      = {American Economic Review},
  year         = {1980},
  volume       = {70},
  number       = {4},
  pages        = {651--659},
  verification = {Confirmed via secondary listings.}
}

@article{graetz1986tax,
  author       = {Graetz, Michael J. and Reinganum, Jennifer F. and Wilde, Louis L.},
  title        = {The Tax Compliance Game: Toward an Interactive Theory of Law Enforcement},
  journal      = {Journal of Law, Economics, and Organization},
  year         = {1986},
  volume       = {2},
  number       = {1},
  pages        = {1--32},
  verification = {Confirmed via RePEc, Caltech repository (published PDF), JSTOR 764914. Table 2 and text: a higher penalty rate lowers both equilibrium auditing and noncompliance.}
}

@article{walker2001minimax,
  author       = {Walker, Mark and Wooders, John},
  title        = {Minimax Play at {Wimbledon}},
  journal      = {American Economic Review},
  year         = {2001},
  volume       = {91},
  number       = {5},
  pages        = {1521--1538},
  doi          = {10.1257/aer.91.5.1521},
  verification = {Confirmed via RePEc.}
}

@article{chiappori2002testing,
  author       = {Chiappori, Pierre-Andr{\'e} and Levitt, Steven and Groseclose, Timothy},
  title        = {Testing Mixed-Strategy Equilibria When Players Are Heterogeneous: The Case of Penalty Kicks in Soccer},
  journal      = {American Economic Review},
  year         = {2002},
  volume       = {92},
  number       = {4},
  pages        = {1138--1151},
  verification = {Confirmed via RePEc.}
}

@article{kleinberg2021algorithmic,
  author       = {Kleinberg, Jon and Raghavan, Manish},
  title        = {Algorithmic Monoculture and Social Welfare},
  journal      = {Proceedings of the National Academy of Sciences},
  year         = {2021},
  volume       = {118},
  number       = {22},
  pages        = {e2018340118},
  doi          = {10.1073/pnas.2018340118},
  verification = {Vol/issue/article confirmed; DOI follows the PNAS article-number pattern.}
}

@article{shachat2004detect,
  author       = {Shachat, Jason and Swarthout, J. Todd},
  title        = {Do We Detect and Exploit Mixed Strategy Play by Opponents?},
  journal      = {Mathematical Methods of Operations Research},
  year         = {2004},
  volume       = {59},
  number       = {3},
  pages        = {359--373},
  doi          = {10.1007/s001860400354},
  verification = {Confirmed via RePEc/Durham.}
}

@article{li2022predictable,
  author       = {Li, Xiaomin and Camerer, Colin F.},
  title        = {Predictable Effects of Visual Salience in Experimental Decisions and Games},
  journal      = {Quarterly Journal of Economics},
  year         = {2022},
  volume       = {137},
  number       = {3},
  pages        = {1849--1900},
  doi          = {10.1093/qje/qjac025},
  verification = {Confirmed via RePEc (IDEAS), 2026-10-13.}
}

@misc{buchanan2026innate,
  author       = {Buchanan, Joy and Foster, Joshua},
  title        = {The Innate Economic Preferences of Language Models},
  howpublished = {arXiv:2607.26288},
  year         = {2026},
  verification = {Confirmed via arXiv abstract page and PDF, 2026-10-13. Position finding in Section 4 (Table 3 and text), all open-weight models but Gemma 4 favor the first of two identical options.}
}

@article{camerer2004cognitive,
  author       = {Camerer, Colin F. and Ho, Teck-Hua and Chong, Juin-Kuan},
  title        = {A Cognitive Hierarchy Model of Games},
  journal      = {Quarterly Journal of Economics},
  year         = {2004},
  volume       = {119},
  number       = {3},
  pages        = {861--898},
  verification = {Confirmed (RePEc URL encodes v119 i3 p861-898; NTU record).}
}

@article{crawford2013structural,
  author       = {Crawford, Vincent P. and Costa-Gomes, Miguel A. and Iriberri, Nagore},
  title        = {Structural Models of Nonequilibrium Strategic Thinking: Theory, Evidence, and Applications},
  journal      = {Journal of Economic Literature},
  year         = {2013},
  volume       = {51},
  number       = {1},
  pages        = {5--62},
  doi          = {10.1257/jel.51.1.5},
  verification = {Confirmed.}
}

@article{palacioshuerta2003professionals,
  author       = {Palacios-Huerta, Ignacio},
  title        = {Professionals Play Minimax},
  journal      = {Review of Economic Studies},
  year         = {2003},
  volume       = {70},
  number       = {2},
  pages        = {395--415},
  verification = {Confirmed (LSE repository; reference lists).}
}

\clearpage
\appendix
\numberwithin{table}{section}
\numberwithin{figure}{section}
\renewcommand{\theHtable}{\thesection.\arabic{table}}
\renewcommand{\theHfigure}{\thesection.\arabic{figure}}
\section*{Online Appendix}
This appendix supplements Sections~\ref{sec:experiment} and~\ref{sec:results} of the paper. Appendix~\ref{app:presentation} reports choices by action letter and listing order. Appendix~\ref{app:reasoning} lists the patterns behind the keyword codes for reasoning and the codes assigned by reading a sample of game replies. Appendix~\ref{app:validity} reports baseline reply validity and truncation for every model, response mode, and temperature. Appendix~\ref{app:random_tool} gives the protocol and results of the randomizer-access treatment. Appendix~\ref{app:game_more} gives game results that Section~\ref{sec:game_results} of the paper summarizes, namely the tests against equilibrium, the prompt variants, the base prize, and temperature, and documents the exploitability and agreement measures of the conclusion. Appendix~\ref{app:history} reports the effects of choice histories in the stated-strategy task. All tables are computed from the raw replies in the replication materials.

\section{Choice Shares by Letter and Listing Order}\label{app:presentation}

The 12 variants that counterbalance presentation cross three letter pairs, J and F, K and Q, and W and Z, with two assignments of letters to actions and two listing orders (Section~\ref{sec:subjects} of the paper). Within a role, a variant is therefore identified by the letter that labels Top or Left and by the position in which that letter is listed. The opponent's letters come from a different pair and are assigned and ordered in parallel with the model's own.

Tables~\ref{tab:variants_nocot} to~\ref{tab:variants_cot_b} report the share choosing Top as Row and Left as Column in each variant and game cell at $u=\$100$ and $T=1$. Each panel ends with the pooled share across the 12 variants at $u=\$100$ (Figure~\ref{fig:game_results} of the paper pools this base prize with $u=\$10$ and $u=\$1{,}000$), the $p$-value of the chi-square test that the 12 variants share a common choice probability, and the number of valid replies. At $k=2$, the game is unchanged when both players' actions are exchanged. The row for a letter listed first and the row for its partner listed second then describe the same prompt, so their shares sum to 100 up to sampling error.

Table~\ref{tab:presentation} summarizes listing-order and letter effects across all cells at $T=1$. Every letter labels each action and appears in each position equally often, so each entry equals 50 percent in the absence of the corresponding effect. In the game, the share of immediate answers that choose the first-listed letter is 58.7 percent for Qwen3.8 27B, 78.4 percent for Gemma~4 31B, and 49.9 percent for Nemotron~3.5 Lightning, whose choices depend instead on the letters, with a share of W 24 points below and of Z 24 points above that of F.

Table~\ref{tab:task_order} reports the stated-strategy task by listing order at each prescribed probability. Listing order matters most at $\pi=1/2$, where Figure~\ref{fig:diagnostic_results} of the paper pools the two orders. Gemini~3.8 Flash chooses the named letter in 35 percent of replies when it is listed first and in 67 percent when it is listed second.

\begin{table}[p]
\centering
\footnotesize
\renewcommand{\arraystretch}{0.95}
\caption{Game Choices by Letter and Listing Order, Immediate Answers, Open-Weight Models}\label{tab:variants_nocot}
\begin{tabular*}{\textwidth}{@{\extracolsep{\fill}}llrrrrrr}
\toprule
& & \multicolumn{3}{c}{Row, share choosing Top} & \multicolumn{3}{c}{Column, share choosing Left} \\
\cmidrule(lr){3-5}\cmidrule(lr){6-8}
Letter & Listed & $k=2$ & $k=4$ & $k=8$ & $k=2$ & $k=4$ & $k=8$ \\
\midrule
\multicolumn{8}{l}{\textit{Panel A. Qwen3.8 27B, immediate answers}} \\
J & First & 77.5 & 45.0 & 45.0 & 45.0 & 40.0 & 22.5 \\
J & Second & 0.0 & 55.0 & 57.5 & 15.0 & 40.0 & 52.5 \\
F & First & 90.0 & 77.5 & 87.5 & 100.0 & 75.0 & 80.0 \\
F & Second & 22.5 & 87.5 & 85.0 & 52.5 & 70.0 & 55.0 \\
K & First & 85.0 & 77.5 & 77.5 & 65.0 & 70.0 & 72.5 \\
K & Second & 12.5 & 92.5 & 95.0 & 40.0 & 67.5 & 97.5 \\
Q & First & 87.5 & 55.0 & 35.0 & 57.5 & 37.5 & 32.5 \\
Q & Second & 12.5 & 50.0 & 55.0 & 22.5 & 47.5 & 37.5 \\
W & First & 90.0 & 77.5 & 70.0 & 70.0 & 55.0 & 47.5 \\
W & Second & 15.0 & 60.0 & 72.5 & 25.0 & 72.5 & 67.5 \\
Z & First & 85.0 & 65.0 & 60.0 & 60.0 & 47.5 & 50.0 \\
Z & Second & 7.5 & 65.0 & 67.5 & 15.0 & 47.5 & 40.0 \\
\addlinespace[2pt]
\multicolumn{2}{l}{All variants} & 48.8 & 67.3 & 67.3 & 47.3 & 55.8 & 54.6 \\
\multicolumn{2}{l}{Common-probability $p$} & $<$0.001 & $<$0.001 & $<$0.001 & $<$0.001 & $<$0.001 & $<$0.001 \\
\addlinespace
\multicolumn{8}{l}{\textit{Panel B. Gemma 4 31B, immediate answers}} \\
J & First & 100.0 & 100.0 & 100.0 & 100.0 & 97.5 & 75.0 \\
J & Second & 0.0 & 0.0 & 100.0 & 2.5 & 2.5 & 12.5 \\
F & First & 100.0 & 100.0 & 100.0 & 100.0 & 62.5 & 2.5 \\
F & Second & 0.0 & 0.0 & 17.5 & 0.0 & 0.0 & 10.0 \\
K & First & 100.0 & 100.0 & 100.0 & 100.0 & 77.5 & 7.5 \\
K & Second & 0.0 & 0.0 & 100.0 & 0.0 & 0.0 & 12.5 \\
Q & First & 100.0 & 100.0 & 100.0 & 100.0 & 92.5 & 97.5 \\
Q & Second & 0.0 & 0.0 & 100.0 & 0.0 & 0.0 & 97.5 \\
W & First & 100.0 & 100.0 & 100.0 & 100.0 & 37.5 & 37.5 \\
W & Second & 0.0 & 0.0 & 100.0 & 0.0 & 5.0 & 97.5 \\
Z & First & 100.0 & 100.0 & 100.0 & 97.5 & 95.0 & 20.0 \\
Z & Second & 0.0 & 0.0 & 90.0 & 0.0 & 0.0 & 7.5 \\
\addlinespace[2pt]
\multicolumn{2}{l}{All variants} & 50.0 & 50.0 & 92.3 & 50.0 & 39.2 & 39.8 \\
\multicolumn{2}{l}{Common-probability $p$} & $<$0.001 & $<$0.001 & $<$0.001 & $<$0.001 & $<$0.001 & $<$0.001 \\
\bottomrule
\end{tabular*}
\begin{minipage}{\textwidth}
\vspace{0.4em}\footnotesize \textit{Notes.} Base prize $u=\$100$ and $T=1$. Each row is one of the 12 variants. ``Letter'' is the letter that labels Top in the Row columns and Left in the Column columns, and ``Listed'' gives its position in the prompt. Entries are percentages of the variant's 40 valid replies. ``All variants'' weights the 12 variants equally. ``Common-probability $p$'' is the $p$-value of a chi-square test that the 12 variants share one choice probability, and dashes mark cells in which every reply chooses the same action. At $k=2$, the row for a letter listed first and the row for its partner listed second present identical prompts, so their shares sum to 100 up to sampling error and these cells measure only responses to presentation.
\end{minipage}
\end{table}

\begin{table}[p]
\centering
\footnotesize
\renewcommand{\arraystretch}{0.95}
\caption{Game Choices by Letter and Listing Order, Immediate Answers, Nemotron~3.5 Lightning}\label{tab:variants_nocot_b}
\begin{tabular*}{\textwidth}{@{\extracolsep{\fill}}llrrrrrr}
\toprule
& & \multicolumn{3}{c}{Row, share choosing Top} & \multicolumn{3}{c}{Column, share choosing Left} \\
\cmidrule(lr){3-5}\cmidrule(lr){6-8}
Letter & Listed & $k=2$ & $k=4$ & $k=8$ & $k=2$ & $k=4$ & $k=8$ \\
\midrule
\multicolumn{8}{l}{\textit{Panel A. Nemotron 3.5 Lightning, immediate answers}} \\
J & First & 52.5 & 52.5 & 50.0 & 40.0 & 45.0 & 60.0 \\
J & Second & 40.0 & 57.5 & 60.0 & 57.5 & 42.5 & 45.0 \\
F & First & 60.0 & 67.5 & 47.5 & 45.0 & 50.0 & 47.5 \\
F & Second & 50.0 & 52.5 & 50.0 & 50.0 & 62.5 & 62.5 \\
K & First & 67.5 & 55.0 & 40.0 & 57.5 & 50.0 & 47.5 \\
K & Second & 55.0 & 75.0 & 72.5 & 57.5 & 77.5 & 57.5 \\
Q & First & 47.5 & 30.0 & 37.5 & 37.5 & 42.5 & 40.0 \\
Q & Second & 47.5 & 50.0 & 52.5 & 52.5 & 55.0 & 55.0 \\
W & First & 27.5 & 32.5 & 35.0 & 27.5 & 22.5 & 17.5 \\
W & Second & 17.5 & 27.5 & 45.0 & 17.5 & 30.0 & 7.5 \\
Z & First & 82.5 & 80.0 & 67.5 & 92.5 & 70.0 & 75.0 \\
Z & Second & 72.5 & 65.0 & 80.0 & 70.0 & 82.5 & 80.0 \\
\addlinespace[2pt]
\multicolumn{2}{l}{All variants} & 51.7 & 53.8 & 53.1 & 50.4 & 52.5 & 49.6 \\
\multicolumn{2}{l}{Common-probability $p$} & $<$0.001 & $<$0.001 & $<$0.001 & $<$0.001 & $<$0.001 & $<$0.001 \\
\bottomrule
\end{tabular*}
\begin{minipage}{\textwidth}
\vspace{0.4em}\footnotesize \textit{Notes.} Base prize $u=\$100$ and $T=1$. Entries are percentages of each variant's 40 valid replies. Rows and summary lines are defined as in Table~\ref{tab:variants_nocot}.
\end{minipage}
\end{table}

\begin{table}[p]
\centering
\footnotesize
\renewcommand{\arraystretch}{0.95}
\caption{Game Choices by Letter and Listing Order, Reasoning, Open-Weight Models}\label{tab:variants_cot}
\begin{tabular*}{\textwidth}{@{\extracolsep{\fill}}llrrrrrr}
\toprule
& & \multicolumn{3}{c}{Row, share choosing Top} & \multicolumn{3}{c}{Column, share choosing Left} \\
\cmidrule(lr){3-5}\cmidrule(lr){6-8}
Letter & Listed & $k=2$ & $k=4$ & $k=8$ & $k=2$ & $k=4$ & $k=8$ \\
\midrule
\multicolumn{8}{l}{\textit{Panel A. Qwen3.8 27B, reasoning}} \\
J & First & 100.0 & 97.5 & 100.0 & 47.5 & 5.0 & 0.0 \\
J & Second & 0.0 & 100.0 & 100.0 & 17.5 & 2.5 & 2.5 \\
F & First & 97.5 & 100.0 & 100.0 & 75.0 & 5.0 & 0.0 \\
F & Second & 0.0 & 100.0 & 100.0 & 50.0 & 0.0 & 2.5 \\
K & First & 100.0 & 100.0 & 100.0 & 75.0 & 0.0 & 5.0 \\
K & Second & 2.5 & 100.0 & 100.0 & 27.5 & 2.5 & 0.0 \\
Q & First & 100.0 & 100.0 & 97.5 & 67.5 & 0.0 & 0.0 \\
Q & Second & 2.5 & 97.5 & 100.0 & 40.0 & 0.0 & 5.0 \\
W & First & 100.0 & 100.0 & 97.5 & 80.0 & 0.0 & 0.0 \\
W & Second & 2.5 & 100.0 & 100.0 & 55.0 & 0.0 & 0.0 \\
Z & First & 100.0 & 100.0 & 100.0 & 55.0 & 0.0 & 0.0 \\
Z & Second & 0.0 & 100.0 & 100.0 & 25.0 & 2.5 & 0.0 \\
\addlinespace[2pt]
\multicolumn{2}{l}{All variants} & 50.4 & 99.6 & 99.6 & 51.2 & 1.5 & 1.3 \\
\multicolumn{2}{l}{Common-probability $p$} & $<$0.001 & 0.527 & 0.527 & $<$0.001 & 0.361 & 0.223 \\
\addlinespace
\multicolumn{8}{l}{\textit{Panel B. Gemma 4 31B, reasoning}} \\
J & First & 100.0 & 97.5 & 97.5 & 95.0 & 25.0 & 20.0 \\
J & Second & 0.0 & 100.0 & 100.0 & 20.0 & 10.0 & 17.5 \\
F & First & 100.0 & 100.0 & 100.0 & 90.0 & 25.0 & 12.5 \\
F & Second & 0.0 & 100.0 & 100.0 & 10.0 & 22.5 & 22.5 \\
K & First & 100.0 & 100.0 & 100.0 & 75.0 & 12.5 & 17.5 \\
K & Second & 0.0 & 100.0 & 97.5 & 2.5 & 27.5 & 25.0 \\
Q & First & 100.0 & 100.0 & 100.0 & 95.0 & 22.5 & 20.0 \\
Q & Second & 0.0 & 100.0 & 100.0 & 17.5 & 12.5 & 7.5 \\
W & First & 100.0 & 100.0 & 100.0 & 87.5 & 17.5 & 15.0 \\
W & Second & 0.0 & 100.0 & 100.0 & 10.0 & 15.0 & 17.5 \\
Z & First & 100.0 & 100.0 & 100.0 & 92.5 & 17.5 & 27.5 \\
Z & Second & 0.0 & 100.0 & 100.0 & 10.0 & 22.5 & 10.0 \\
\addlinespace[2pt]
\multicolumn{2}{l}{All variants} & 50.0 & 99.8 & 99.6 & 50.4 & 19.2 & 17.7 \\
\multicolumn{2}{l}{Common-probability $p$} & $<$0.001 & 0.441 & 0.527 & $<$0.001 & 0.579 & 0.490 \\
\addlinespace
\multicolumn{8}{l}{\textit{Panel C. Nemotron 3.5 Lightning, reasoning}} \\
J & First & 100.0 & 100.0 & 100.0 & 100.0 & 95.0 & 97.5 \\
J & Second & 0.0 & 87.5 & 92.5 & 0.0 & 2.5 & 2.5 \\
F & First & 100.0 & 100.0 & 100.0 & 100.0 & 95.0 & 87.5 \\
F & Second & 0.0 & 95.0 & 92.5 & 0.0 & 2.5 & 5.0 \\
K & First & 100.0 & 100.0 & 100.0 & 100.0 & 85.0 & 90.0 \\
K & Second & 2.5 & 87.5 & 82.5 & 0.0 & 2.5 & 0.0 \\
Q & First & 100.0 & 100.0 & 100.0 & 100.0 & 92.5 & 92.5 \\
Q & Second & 0.0 & 100.0 & 100.0 & 0.0 & 7.5 & 5.0 \\
W & First & 100.0 & 100.0 & 100.0 & 100.0 & 85.0 & 82.5 \\
W & Second & 0.0 & 80.0 & 92.5 & 0.0 & 0.0 & 0.0 \\
Z & First & 100.0 & 100.0 & 100.0 & 100.0 & 100.0 & 95.0 \\
Z & Second & 0.0 & 97.5 & 100.0 & 0.0 & 5.0 & 2.5 \\
\addlinespace[2pt]
\multicolumn{2}{l}{All variants} & 50.2 & 95.6 & 96.7 & 50.0 & 47.7 & 46.7 \\
\multicolumn{2}{l}{Common-probability $p$} & $<$0.001 & $<$0.001 & $<$0.001 & $<$0.001 & $<$0.001 & $<$0.001 \\
\bottomrule
\end{tabular*}
\begin{minipage}{\textwidth}
\vspace{0.4em}\footnotesize \textit{Notes.} Base prize $u=\$100$ and $T=1$. Entries are percentages of each variant's valid replies, 40 per variant. Rows and summary lines are defined as in Table~\ref{tab:variants_nocot}.
\end{minipage}
\end{table}

\begin{table}[p]
\centering
\footnotesize
\renewcommand{\arraystretch}{0.95}
\caption{Game Choices by Letter and Listing Order, Reasoning, Frontier Models}\label{tab:variants_cot_b}
\begin{tabular*}{\textwidth}{@{\extracolsep{\fill}}llrrrrrr}
\toprule
& & \multicolumn{3}{c}{Row, share choosing Top} & \multicolumn{3}{c}{Column, share choosing Left} \\
\cmidrule(lr){3-5}\cmidrule(lr){6-8}
Letter & Listed & $k=2$ & $k=4$ & $k=8$ & $k=2$ & $k=4$ & $k=8$ \\
\midrule
\multicolumn{8}{l}{\textit{Panel A. Gemini 3.8 Flash, reasoning}} \\
J & First & 100 & 80 & 100 & 70 & 0 & 0 \\
J & Second & 0 & 40 & 90 & 20 & 0 & 0 \\
F & First & 100 & 80 & 100 & 80 & 0 & 0 \\
F & Second & 0 & 40 & 100 & 60 & 0 & 0 \\
K & First & 90 & 60 & 100 & 60 & 0 & 0 \\
K & Second & 0 & 50 & 100 & 60 & 0 & 0 \\
Q & First & 100 & 80 & 100 & 70 & 0 & 0 \\
Q & Second & 0 & 60 & 90 & 40 & 0 & 0 \\
W & First & 100 & 80 & 100 & 60 & 0 & 0 \\
W & Second & 0 & 60 & 100 & 50 & 0 & 0 \\
Z & First & 100 & 60 & 100 & 40 & 0 & 0 \\
Z & Second & 0 & 50 & 100 & 50 & 0 & 0 \\
\addlinespace[2pt]
\multicolumn{2}{l}{All variants} & 49 & 62 & 98 & 55 & 0 & 0 \\
\multicolumn{2}{l}{Common-probability $p$} & $<$0.001 & 0.455 & 0.515 & 0.385 & -- & -- \\
\addlinespace
\multicolumn{8}{l}{\textit{Panel B. Claude Haiku 5.5, reasoning}} \\
J & First & 100 & 100 & 90 & 100 & 40 & 20 \\
J & Second & 0 & 100 & 90 & 0 & 20 & 0 \\
F & First & 100 & 100 & 80 & 100 & 10 & 10 \\
F & Second & 0 & 90 & 90 & 0 & 10 & 10 \\
K & First & 100 & 100 & 90 & 100 & 20 & 10 \\
K & Second & 0 & 100 & 100 & 30 & 10 & 50 \\
Q & First & 100 & 100 & 90 & 90 & 20 & 30 \\
Q & Second & 0 & 100 & 100 & 0 & 20 & 10 \\
W & First & 100 & 100 & 100 & 80 & 20 & 10 \\
W & Second & 0 & 90 & 80 & 30 & 10 & 40 \\
Z & First & 90 & 100 & 100 & 90 & 40 & 30 \\
Z & Second & 0 & 90 & 100 & 0 & 10 & 10 \\
\addlinespace[2pt]
\multicolumn{2}{l}{All variants} & 49 & 98 & 93 & 52 & 19 & 19 \\
\multicolumn{2}{l}{Common-probability $p$} & $<$0.001 & 0.601 & 0.621 & $<$0.001 & 0.683 & 0.138 \\
\addlinespace
\multicolumn{8}{l}{\textit{Panel C. GPT-6 Luna, reasoning}} \\
J & First & 100 & 100 & 100 & 60 & 20 & 20 \\
J & Second & 0 & 100 & 100 & 10 & 30 & 30 \\
F & First & 100 & 90 & 90 & 100 & 60 & 30 \\
F & Second & 0 & 100 & 100 & 10 & 20 & 0 \\
K & First & 100 & 100 & 100 & 90 & 60 & 30 \\
K & Second & 10 & 100 & 90 & 10 & 50 & 50 \\
Q & First & 100 & 90 & 100 & 80 & 30 & 30 \\
Q & Second & 0 & 80 & 100 & 0 & 20 & 20 \\
W & First & 100 & 100 & 100 & 90 & 20 & 20 \\
W & Second & 0 & 100 & 90 & 30 & 50 & 40 \\
Z & First & 90 & 100 & 100 & 70 & 40 & 50 \\
Z & Second & 0 & 90 & 100 & 30 & 60 & 70 \\
\addlinespace[2pt]
\multicolumn{2}{l}{All variants} & 50 & 96 & 98 & 48 & 38 & 32 \\
\multicolumn{2}{l}{Common-probability $p$} & $<$0.001 & 0.341 & 0.601 & $<$0.001 & 0.268 & 0.123 \\
\bottomrule
\end{tabular*}
\begin{minipage}{\textwidth}
\vspace{0.4em}\footnotesize \textit{Notes.} Base prize $u=\$100$. Gemini~3.8 Flash and GPT-6 Luna run at $T=1$, and Claude Haiku~5.5 at its default temperature. Entries are percentages of each variant's 10 valid replies. Rows and summary lines are defined as in Table~\ref{tab:variants_nocot}.
\end{minipage}
\end{table}

\begin{table}[p]
\centering
\footnotesize
\caption{Listing-Order and Letter Effects at $T=1$}\label{tab:presentation}
\begin{tabular*}{\textwidth}{@{\extracolsep{\fill}}llrrrrrrrr}
\toprule
& & \multicolumn{4}{c}{Game} & \multicolumn{4}{c}{Stated-strategy task} \\
\cmidrule(lr){3-6}\cmidrule(lr){7-10}
Model & Mode & First & J & K & W & First & J & K & W \\
\midrule
Qwen3.8 27B & Immediate & 58.7 & 35.5 & 64.7 & 52.4 & 52.2 & 54.2 & 56.1 & 50.7 \\
Gemma 4 31B & Immediate & 78.4 & 54.1 & 38.4 & 54.6 & 47.6 & 48.8 & 48.6 & 47.8 \\
Nemotron 3.5 Lightning & Immediate & 49.9 & 49.7 & 56.2 & 25.9 & 36.0 & 52.4 & 52.4 & 38.5 \\
Qwen3.8 27B & Reasoning & 60.9 & 47.6 & 49.9 & 52.8 & 49.9 & 49.7 & 50.2 & 50.4 \\
Gemma 4 31B & Reasoning & 65.7 & 50.1 & 49.5 & 48.9 & 47.3 & 49.7 & 49.8 & 48.9 \\
Nemotron 3.5 Lightning & Reasoning & 81.9 & 50.0 & 48.7 & 47.5 & 49.1 & 50.1 & 49.6 & 46.4 \\
Claude Haiku 5.5 & Reasoning & 67 & 50 & 50 & 48 & 50 & 51 & 53 & 51 \\
GPT-6 Luna & Reasoning & 61 & 49 & 53 & 51 & 49 & 49 & 50 & 49 \\
Gemini 3.8 Flash & Reasoning & 61 & 48 & 48 & 52 & 48 & 49 & 50 & 51 \\
\bottomrule
\end{tabular*}
\begin{minipage}{\textwidth}
\vspace{0.4em}\footnotesize \textit{Notes.} Entries are percentages of valid replies at $T=1$. ``First'' is the share choosing the letter listed first. ``J'' is the share choosing J among replies to prompts that offer J and F, and ``K'' and ``W'' are defined in the same way for the pairs K and Q and W and Z. Each letter is assigned to each action, and listed in each position, in an equal number of draws, so every entry equals 50 in the absence of a listing-order or letter effect. Game entries pool the 18 game cells for the open-weight models and the six cells with $u=\$100$ for the frontier models. Task entries pool the nine prescribed probabilities. Dashes mark arms that were not run.
\end{minipage}
\end{table}

\begin{table}[p]
\centering
\scriptsize
\setlength{\tabcolsep}{3pt}
\caption{Stated-Strategy Task, Share Choosing the Named Letter by Listing Order at $T=1$}\label{tab:task_order}
\begin{tabular*}{\textwidth}{@{\extracolsep{\fill}}lllrrrrrrrrr}
\toprule
& & & \multicolumn{9}{c}{Prescribed probability of the named letter, $\pi$} \\
\cmidrule(lr){4-12}
Model & Mode & Named letter & $1/100$ & $1/10$ & $1/4$ & $1/3$ & $1/2$ & $2/3$ & $3/4$ & $9/10$ & $99/100$ \\
\midrule
\multirow{2}{*}{Qwen3.8 27B} & \multirow{2}{*}{Immediate} & First & 20.8 & 9.6 & 19.6 & 33.3 & 77.1 & 92.9 & 89.2 & 92.9 & 96.2 \\
&  & Second & 15.8 & 14.2 & 21.7 & 28.3 & 55.8 & 84.2 & 89.6 & 85.8 & 97.5 \\
\addlinespace
\multirow{2}{*}{Gemma 4 31B} & \multirow{2}{*}{Immediate} & First & 0.0 & 0.0 & 0.0 & 0.0 & 53.3 & 100.0 & 100.0 & 100.0 & 100.0 \\
&  & Second & 0.0 & 0.0 & 0.0 & 0.0 & 97.1 & 100.0 & 100.0 & 100.0 & 100.0 \\
\addlinespace
\multirow{2}{*}{Nemotron 3.5 Lightning} & \multirow{2}{*}{Immediate} & First & 37.5 & 46.7 & 47.9 & 53.8 & 66.2 & 78.8 & 80.4 & 82.1 & 79.2 \\
&  & Second & 73.8 & 82.5 & 89.6 & 87.5 & 95.8 & 98.3 & 98.3 & 98.8 & 99.2 \\
\addlinespace
\multirow{2}{*}{Qwen3.8 27B} & \multirow{2}{*}{Reasoning} & First & 0.0 & 0.0 & 0.0 & 0.0 & 75.4 & 100.0 & 100.0 & 100.0 & 100.0 \\
&  & Second & 0.0 & 0.0 & 0.0 & 0.0 & 77.9 & 100.0 & 100.0 & 100.0 & 100.0 \\
\addlinespace
\multirow{2}{*}{Gemma 4 31B} & \multirow{2}{*}{Reasoning} & First & 0.0 & 0.0 & 0.0 & 0.0 & 45.4 & 100.0 & 100.0 & 100.0 & 100.0 \\
&  & Second & 0.0 & 0.0 & 0.0 & 0.0 & 94.6 & 100.0 & 100.0 & 100.0 & 100.0 \\
\addlinespace
\multirow{2}{*}{Nemotron 3.5 Lightning} & \multirow{2}{*}{Reasoning} & First & 0.4 & 1.7 & 0.4 & 3.3 & 84.2 & 97.9 & 100.0 & 100.0 & 99.6 \\
&  & Second & 0.0 & 2.5 & 3.3 & 4.2 & 95.0 & 99.6 & 100.0 & 100.0 & 99.6 \\
\addlinespace
\multirow{2}{*}{Claude Haiku 5.5} & \multirow{2}{*}{Reasoning} & First & 0 & 0 & 0 & 0 & 67 & 100 & 100 & 100 & 100 \\
&  & Second & 0 & 0 & 0 & 0 & 58 & 100 & 100 & 100 & 100 \\
\addlinespace
\multirow{2}{*}{GPT-6 Luna} & \multirow{2}{*}{Reasoning} & First & 0 & 0 & 0 & 2 & 77 & 100 & 100 & 100 & 100 \\
&  & Second & 0 & 0 & 3 & 5 & 92 & 100 & 100 & 100 & 100 \\
\addlinespace
\multirow{2}{*}{Gemini 3.8 Flash} & \multirow{2}{*}{Reasoning} & First & 0 & 0 & 0 & 0 & 35 & 100 & 100 & 100 & 100 \\
&  & Second & 0 & 0 & 0 & 0 & 67 & 100 & 100 & 100 & 100 \\
\bottomrule
\end{tabular*}
\begin{minipage}{\textwidth}
\vspace{0.4em}\footnotesize \textit{Notes.} Entries are percentages of valid replies that choose the letter named in the instruction (Figure~1, Panel~B), separately for prompts that list the named letter first and second. Each entry rests on 912 valid replies for the open-weight models and 60 for the frontier models. Claude Haiku~5.5 runs at its default temperature.
\end{minipage}
\end{table}

\clearpage
\section{Reasoning Codes}\label{app:reasoning}

Section~\ref{sec:game_results} of the paper describes the reasoning behind game choices with two coding methods, keyword codes applied to every reply and codes assigned by reading a seeded random sample. Both were defined after I saw the data. Section~\ref{sec:diagnostic} applies keyword codes to the stated-strategy task in the same way.

\subsection{Keyword Codes}

Table~\ref{tab:patterns} lists each code and its pattern. A reply carries a code when the pattern matches anywhere in its stored reasoning, without regard to case. For the open-weight models, the stored reasoning is the full trace that precedes the visible answer. For the frontier models, it is the provider's summary of the reasoning followed by the visible reply, so their codes measure what the summary reports. Keyword codes measure mentions rather than use. A reply that names a mixed strategy need not play one, and the pattern for joint earnings matches any use of the words combined, joint, or total.

Table~\ref{tab:game_codes} reports the share of game replies carrying each code, by model, role, and $k$, at $u=\$100$. Table~\ref{tab:task_codes} reports the codes in the stated-strategy task.

\begin{table}[p]
\centering
\footnotesize
\caption{Keyword Patterns for Reasoning Codes}\label{tab:patterns}
\begin{tabularx}{\textwidth}{>{\raggedright\arraybackslash}p{0.17\textwidth}>{\raggedright\arraybackslash}p{0.30\textwidth}>{\raggedright\arraybackslash}X}
\toprule
Code & Content & Pattern \\
\midrule
\multicolumn{3}{l}{\textit{Panel A. Game}} \\
Nash & Names the Nash equilibrium & \texttt{nash} \\
Equilibrium & Uses the word equilibrium in any form & \texttt{equilibri} \\
Mixed strategy & Names a mixed strategy & \texttt{mixed[-\textvisiblespace{}]strateg} \\
Arbitrary & Calls the choice arbitrary & \texttt{arbitrar} \\
Random device & Mentions a randomizing device or a random draw & \texttt{\textbackslash{}bdie\textbackslash{}b\textbar{}\allowbreak{}\textbackslash{}bdice\textbackslash{}b\textbar{}\allowbreak{}\textbackslash{}bcoin\textbackslash{}b\textbar{}\allowbreak{}\textbackslash{}broll\textbar{}\allowbreak{}random\textvisiblespace{}number\textbar{}\allowbreak{}generate\textvisiblespace{}a\textvisiblespace{}(?:random\textvisiblespace{})?number\textbar{}\allowbreak{}pick\textvisiblespace{}a\textvisiblespace{}number\textbar{}\allowbreak{}spinner\textbar{}\allowbreak{}random(?:ly)?\textvisiblespace{}(?:select\textbar{}\allowbreak{}pick\textbar{}\allowbreak{}choos)} \\
Coin, die, or flip & Mentions a coin, a die, or a flip & \texttt{\textbackslash{}bcoins?\textbackslash{}b\textbar{}\allowbreak{}\textbackslash{}bdie\textbackslash{}b\textbar{}\allowbreak{}\textbackslash{}bdice\textbackslash{}b\textbar{}\allowbreak{}\textbackslash{}bflip} \\
Joint earnings & Refers to combined, joint, or total earnings & \texttt{combined\textbar{}\allowbreak{}joint\textbar{}\allowbreak{}total} \\
Goeree and Holt & Names Goeree or Holt & \texttt{goeree\textbar{}\allowbreak{}holt} \\
\addlinespace
\multicolumn{3}{l}{\textit{Panel B. Stated-strategy task}} \\
Random device & Mentions a randomizing device or a random draw & \texttt{\textbackslash{}bdie\textbackslash{}b\textbar{}\allowbreak{}\textbackslash{}bdice\textbackslash{}b\textbar{}\allowbreak{}\textbackslash{}bcoin\textbackslash{}b\textbar{}\allowbreak{}\textbackslash{}broll\textbar{}\allowbreak{}random\textvisiblespace{}number\textbar{}\allowbreak{}generate\textvisiblespace{}a\textvisiblespace{}(?:random\textvisiblespace{})?number\textbar{}\allowbreak{}pick\textvisiblespace{}a\textvisiblespace{}number\textbar{}\allowbreak{}spinner\textbar{}\allowbreak{}random(?:ly)?\textvisiblespace{}(?:select\textbar{}\allowbreak{}pick\textbar{}\allowbreak{}choos)} \\
Cannot randomize & States that the model cannot generate a random outcome & \texttt{can(?:'\textbar{}\allowbreak{}no)t\textvisiblespace{}(?:actually\textvisiblespace{}\textbar{}\allowbreak{}truly\textvisiblespace{}\textbar{}\allowbreak{}really\textvisiblespace{})?(?:generate\textbar{}\allowbreak{}produce\textbar{}\allowbreak{}be)\textvisiblespace{}(?:a\textvisiblespace{})?random\textbar{}\allowbreak{}can(?:'\textbar{}\allowbreak{}no)t\textvisiblespace{}(?:actually\textvisiblespace{}\textbar{}\allowbreak{}truly\textvisiblespace{}\textbar{}\allowbreak{}really\textvisiblespace{})?randomi\textbar{}\allowbreak{}unable\textvisiblespace{}to\textvisiblespace{}(?:generate\textbar{}\allowbreak{}randomi)\textbar{}\allowbreak{}no\textvisiblespace{}(?:actual\textbar{}\allowbreak{}real\textbar{}\allowbreak{}true)\textvisiblespace{}random\textbar{}\allowbreak{}as\textvisiblespace{}an\textvisiblespace{}ai} \\
Imagined outcome & States the outcome of an imagined draw & \texttt{lands\textvisiblespace{}on\textbar{}\allowbreak{}landed\textvisiblespace{}on\textbar{}\allowbreak{}rolled\textvisiblespace{}an?\textvisiblespace{}\textbackslash{}w\textbar{}\allowbreak{}the\textvisiblespace{}(?:random\textvisiblespace{})?number\textvisiblespace{}(?:is\textbar{}\allowbreak{}was)\textvisiblespace{}\textbackslash{}d\textbar{}\allowbreak{}(?:result\textbar{}\allowbreak{}outcome)\textvisiblespace{}(?:is\textbar{}\allowbreak{}was)\textvisiblespace{}(?:a\textvisiblespace{}\textbar{}\allowbreak{}an\textvisiblespace{})?\textbackslash{}d\textbar{}\allowbreak{}i\textvisiblespace{}(?:got\textbar{}\allowbreak{}get)\textvisiblespace{}(?:a\textvisiblespace{}\textbar{}\allowbreak{}an\textvisiblespace{})?\textbackslash{}d\textbar{}\allowbreak{}assume\textvisiblespace{}(?:that\textvisiblespace{})?the\textvisiblespace{}(?:random\textbar{}\allowbreak{}outcome\textbar{}\allowbreak{}result\textbar{}\allowbreak{}roll)} \\
\bottomrule
\end{tabularx}
\begin{minipage}{\textwidth}
\vspace{0.4em}\footnotesize \textit{Notes.} Patterns are Python regular expressions, matched without regard to case against the stored reasoning. For the frontier models this is the provider's summary of the reasoning followed by the visible reply. A reply carries a code if the pattern matches anywhere in its reasoning. The symbol \texttt{\textvisiblespace} marks a space, \texttt{\textbackslash b} marks a word boundary, and \texttt{(?:\dots)} groups alternatives. Patterns were chosen after reading replies.
\end{minipage}
\end{table}

\begin{table}[p]
\centering
\footnotesize
\setlength{\tabcolsep}{4pt}\renewcommand{\arraystretch}{0.9}
\caption{Reasoning Codes in the Game, Share of Replies}\label{tab:game_codes}
\begin{tabular*}{\textwidth}{@{\extracolsep{\fill}}llrrrrrrrrr}
\toprule
Model & Cell & Replies & \rotatebox{90}{Nash} & \rotatebox{90}{Equilibrium} & \rotatebox{90}{Mixed strategy} & \rotatebox{90}{Arbitrary} & \rotatebox{90}{Random device} & \rotatebox{90}{Coin, die, or flip} & \rotatebox{90}{Joint earnings} & \rotatebox{90}{Goeree and Holt} \\
\midrule
Qwen3.8 27B & Row, $k=2$ & 480 & 95.6 & 98.1 & 74.6 & 98.5 & 33.5 & 32.7 & 62.3 & 0.0 \\
 & Row, $k=4$ & 480 & 99.6 & 100.0 & 93.1 & 76.0 & 35.6 & 29.8 & 90.6 & 0.0 \\
 & Row, $k=8$ & 480 & 99.8 & 100.0 & 91.7 & 70.4 & 34.8 & 27.9 & 86.7 & 0.0 \\
 & Column, $k=2$ & 480 & 98.3 & 99.4 & 90.6 & 100.0 & 61.9 & 58.3 & 79.6 & 0.2 \\
 & Column, $k=4$ & 480 & 100.0 & 100.0 & 98.1 & 87.7 & 40.4 & 31.0 & 97.7 & 0.2 \\
 & Column, $k=8$ & 480 & 100.0 & 100.0 & 97.7 & 82.3 & 37.3 & 27.9 & 97.5 & 0.2 \\
 & All & 2,880 & 98.9 & 99.6 & 91.0 & 85.8 & 40.6 & 34.6 & 85.7 & 0.1 \\
\addlinespace
Gemma 4 31B & Row, $k=2$ & 480 & 96.5 & 97.1 & 87.7 & 41.0 & 23.1 & 18.1 & 84.8 & 0.0 \\
 & Row, $k=4$ & 480 & 99.8 & 100.0 & 99.8 & 2.3 & 38.1 & 32.7 & 77.3 & 0.0 \\
 & Row, $k=8$ & 480 & 99.6 & 99.8 & 97.9 & 1.2 & 32.7 & 27.1 & 47.9 & 0.0 \\
 & Column, $k=2$ & 480 & 97.5 & 97.7 & 90.0 & 42.5 & 30.2 & 26.2 & 89.6 & 0.0 \\
 & Column, $k=4$ & 480 & 99.8 & 99.8 & 99.8 & 1.5 & 19.4 & 14.6 & 96.5 & 0.0 \\
 & Column, $k=8$ & 480 & 99.4 & 99.4 & 96.7 & 0.4 & 11.5 & 7.5 & 85.2 & 0.0 \\
 & All & 2,880 & 98.8 & 99.0 & 95.3 & 14.8 & 25.8 & 21.0 & 80.2 & 0.0 \\
\addlinespace
Nemotron 3.5 Lightning & Row, $k=2$ & 480 & 99.6 & 100.0 & 85.0 & 96.5 & 72.5 & 71.9 & 61.5 & 0.0 \\
 & Row, $k=4$ & 480 & 100.0 & 100.0 & 95.4 & 54.4 & 36.7 & 38.1 & 47.3 & 0.2 \\
 & Row, $k=8$ & 480 & 100.0 & 100.0 & 96.7 & 36.2 & 35.4 & 36.5 & 47.3 & 0.0 \\
 & Column, $k=2$ & 480 & 99.8 & 99.8 & 93.5 & 96.7 & 73.1 & 71.7 & 76.2 & 0.0 \\
 & Column, $k=4$ & 480 & 99.6 & 100.0 & 98.1 & 92.1 & 75.6 & 74.8 & 62.9 & 0.0 \\
 & Column, $k=8$ & 480 & 100.0 & 100.0 & 97.9 & 88.5 & 77.1 & 76.2 & 64.2 & 0.0 \\
 & All & 2,880 & 99.8 & 100.0 & 94.4 & 77.4 & 61.7 & 61.5 & 59.9 & 0.0 \\
\addlinespace
Gemini 3.8 Flash & Row, $k=2$ & 120 & 100 & 100 & 96 & 30 & 6 & 3 & 93 & 0 \\
 & Row, $k=4$ & 120 & 100 & 100 & 100 & 0 & 8 & 5 & 25 & 42 \\
 & Row, $k=8$ & 120 & 100 & 100 & 100 & 0 & 6 & 8 & 34 & 27 \\
 & Column, $k=2$ & 120 & 100 & 100 & 91 & 22 & 6 & 8 & 88 & 0 \\
 & Column, $k=4$ & 120 & 100 & 100 & 100 & 0 & 8 & 9 & 26 & 3 \\
 & Column, $k=8$ & 120 & 100 & 100 & 99 & 0 & 5 & 8 & 33 & 8 \\
 & All & 720 & 100 & 100 & 98 & 9 & 6 & 7 & 50 & 13 \\
\addlinespace
Claude Haiku 5.5 & Row, $k=2$ & 120 & 54 & 100 & 57 & 37 & 29 & 35 & 38 & 0 \\
 & Row, $k=4$ & 120 & 78 & 98 & 85 & 0 & 2 & 11 & 73 & 0 \\
 & Row, $k=8$ & 120 & 85 & 100 & 96 & 0 & 0 & 8 & 67 & 0 \\
 & Column, $k=2$ & 120 & 58 & 98 & 56 & 39 & 26 & 39 & 47 & 0 \\
 & Column, $k=4$ & 120 & 81 & 100 & 91 & 0 & 2 & 18 & 83 & 0 \\
 & Column, $k=8$ & 120 & 88 & 100 & 92 & 0 & 0 & 13 & 79 & 0 \\
 & All & 720 & 74 & 99 & 80 & 13 & 10 & 21 & 65 & 0 \\
\addlinespace
GPT-6 Luna & Row, $k=2$ & 120 & 0 & 12 & 6 & 32 & 3 & 0 & 9 & 0 \\
 & Row, $k=4$ & 120 & 6 & 22 & 13 & 2 & 0 & 0 & 1 & 0 \\
 & Row, $k=8$ & 120 & 7 & 21 & 8 & 1 & 0 & 0 & 1 & 0 \\
 & Column, $k=2$ & 120 & 1 & 6 & 3 & 32 & 1 & 0 & 7 & 0 \\
 & Column, $k=4$ & 120 & 9 & 32 & 22 & 14 & 1 & 0 & 10 & 0 \\
 & Column, $k=8$ & 120 & 11 & 42 & 32 & 13 & 1 & 0 & 14 & 0 \\
 & All & 720 & 6 & 22 & 14 & 16 & 1 & 0 & 7 & 0 \\
\bottomrule
\end{tabular*}
\begin{minipage}{\textwidth}
\vspace{0.4em}\footnotesize \textit{Notes.} Entries are percentages of valid replies whose reasoning matches the pattern in Table~\ref{tab:patterns}, Panel~A. All replies are at $u=\$100$ and $T=1$ (the default temperature for Claude Haiku~5.5).
\end{minipage}
\end{table}

\begin{table}[p]
\centering
\footnotesize
\setlength{\tabcolsep}{3.5pt}
\caption{Reasoning Codes in the Stated-Strategy Task}\label{tab:task_codes}
\begin{tabular*}{\textwidth}{@{\extracolsep{\fill}}lrrrrrr}
\toprule
& & \multicolumn{3}{c}{All targets, share of replies} & \multicolumn{2}{c}{Unequal targets} \\
\cmidrule(lr){3-5}\cmidrule(lr){6-7}
Model & Replies & \makecell[r]{Random\\device} & \makecell[r]{Cannot\\randomize} & \makecell[r]{Imagined\\outcome} & Replies & \makecell[r]{Less likely\\letter} \\
\midrule
Qwen3.8 27B & 4,320 & 69.0 & 67.7 & 16.4 & 3,840 & 0 \\
Gemma 4 31B & 4,320 & 99.7 & 5.2 & 64.5 & 3,840 & 0 \\
Nemotron 3.5 Lightning & 4,320 & 100.0 & 88.1 & 82.2 & 3,840 & 46 \\
Claude Haiku 5.5 & 1,080 & 42 & 45 & 8 & 960 & 0 \\
GPT-6 Luna & 1,080 & 10 & 2 & 0 & 960 & 6 \\
Gemini 3.8 Flash & 1,080 & 99 & 0 & 36 & 960 & 0 \\
\bottomrule
\end{tabular*}
\begin{minipage}{\textwidth}
\vspace{0.4em}\footnotesize \textit{Notes.} Valid reasoning replies at $T=1$ (the default temperature for Claude Haiku~5.5). The three code columns give the percentage of all replies whose reasoning matches the pattern in Table~\ref{tab:patterns}, Panel~B. The last two columns give the number of replies under unequal targets ($\pi\neq 1/2$) and how many of them choose the less likely letter.
\end{minipage}
\end{table}

\clearpage
\subsection{Codes Assigned by Reading}

The reading sample is drawn with a fixed seed (20261012) from valid replies at $T=1$ and $u=\$100$ in the cells with $k=4$ and $k=8$, where the equilibrium prediction for Column departs from one half. It contains 12 Qwen3.8 27B replies, six in each role, and 20 Gemini~3.8 Flash replies, ten in each role. It also contains 16 Gemma~4 31B replies, six as Row and ten as Column, and 20 Claude Haiku~5.5 Column replies. The Column replies of Gemma~4 and Haiku are split evenly between Left and Right, because the Column shares of both models lie near equilibrium and the question is how replies choosing each action reach that choice. Gemma~4's Row replies are drawn without regard to the action chosen, since it chooses Top in nearly every Row reply at $k=4$ and $k=8$.

Each sampled reply was read in full and assigned three codes. The first records how the reply evaluates the game, the second records the reason it gives for its final choice, and the third records any further observation, such as whether it computes the mixed-strategy Nash equilibrium (MSNE) correctly. Table~\ref{tab:reading} groups the 68 replies by the codes they share. The replication materials give the codes for each reply together with its location in the raw data.

The sample supports the statements in Section~\ref{sec:game_results} of the paper as follows. All 12 Qwen3.8 replies compute the MSNE correctly and then choose a single action. Five of the six Row replies choose Top because it earns more when Column's two actions are treated as equally likely and because its largest regret is smaller, and the sixth cites the regret comparison and Top's larger maximum payoff. All six Column replies choose Right and cite its larger equilibrium weight, and five of them also predict that Row is drawn to the large prize. All 16 Gemma~4 replies compute the MSNE correctly. All six Row replies compare the probability of Left with the threshold $1/k$ and choose Top. The five Column replies choosing Right predict that Row is drawn to the large prize and cite the larger equilibrium weight of Right, and the five choosing Left conclude that the model earns \$100 whichever action it chooses and break the tie by joint earnings. All eight Gemini Row replies that choose Top compare the probability of Left with the threshold $1/k$, and all ten Gemini Column replies predict that Row is drawn to the large prize and choose Right. All 20 Haiku replies compute the equilibrium mix. The ten choosing Right select the action with the larger equilibrium weight, and the ten choosing Left break the tie by joint earnings.

Two further patterns appear in the keyword codes and the sample. Gemma~4 mentions a coin, a die, or a flip in 21.0 percent of its reasoning replies at $u=\$100$. Gemini's beliefs differ by role in a way that level-$k$ models of strategic thinking generate \citep{camerer2004cognitive,crawford2013structural}. As Row, it judges that Column will choose Left often enough to make Top pay, and as Column, it predicts that Row will be drawn to the large prize and chooses Right. Against a Column player who always chooses Right, Row earns more from Bottom, so the two beliefs are inconsistent. A level-1 Row player who expects Column to choose each action with probability one half earns $(k+1)u/2$ from Top and $1.5u$ from Bottom, so it chooses Top whenever $k>2$. A level-1 Column player who expects the same of Row is indifferent, and a level-2 Column player, who expects Row to favor Top, chooses Right. Gemini's Row replies thus match level~1 and its Column replies level~2. Qwen3.8's sampled replies make the same two steps explicit. Five of its six Row replies treat Column's actions as equally likely, and five of its six Column replies compute Row's expected earnings under the same assumption and conclude that Row favors Top.

\footnotesize
\setlength{\tabcolsep}{4pt}
\begin{longtable}{@{}>{\raggedright\arraybackslash}p{0.11\textwidth}llr>{\raggedright\arraybackslash}p{0.21\textwidth}>{\raggedright\arraybackslash}p{0.19\textwidth}>{\raggedright\arraybackslash}p{0.21\textwidth}@{}}
\caption{Codes Assigned by Reading a Sample of Game Replies}\label{tab:reading} \\
\toprule
Model & Role & Choice & $n$ & How the reply evaluates the game & Reason given for the choice & Other observations \\
\midrule
\endfirsthead
\multicolumn{7}{l}{\textit{Table \thetable, continued}} \\
\toprule
Model & Role & Choice & $n$ & How the reply evaluates the game & Reason given for the choice & Other observations \\
\midrule
\endhead
\bottomrule
\multicolumn{7}{@{}p{\textwidth}@{}}{\rule{0pt}{1.2em}\footnotesize \textit{Notes.} Valid replies at $T=1$ and $u=\$100$ with $k\in\{4,8\}$, drawn with a fixed seed. Each row groups the sampled replies that share a model, role, choice, and the two main codes, and $n$ counts them. Other observations give the number of replies in the group with each note. MSNE denotes the mixed-strategy Nash equilibrium, and a model's own equilibrium mix is its equilibrium probability of Left as Column, $1/k$. The code for each sampled reply, with its location in the raw data, is in the replication materials.} \\
\endlastfoot
Qwen3.8 27B & Row & Top & 5 & Computes the threshold $1/k$ for Left & Treats the opponent's actions as equally likely and minimizes maximum regret & Computes the MSNE correctly (5) \\
 & Row & Top & 1 & Computes the threshold $1/k$ for Left & Minimizes maximum regret and offers the larger maximum payoff & Computes the MSNE correctly (1) \\
 & Column & Right & 5 & Predicts that Row is drawn to the large prize & Chooses Right as the best response to Top and cites the equilibrium weight & States its own equilibrium mix correctly (5) \\
 & Column & Right & 1 & Computes the equilibrium mix & Larger equilibrium weight and the opponent's smaller payoff gap after it & States its own equilibrium mix correctly (1) \\
\addlinespace
Gemma 4 31B & Row & Top & 3 & Computes the threshold $1/k$ for Left & Judges that the probability of Left exceeds $1/k$ & Computes the MSNE correctly (3) \\
 & Row & Top & 3 & Computes the threshold $1/k$ for Left & Top has the same minimum payoff and a larger maximum & Computes the MSNE correctly (3) \\
 & Column & Left & 5 & Opponent best responds to own choice, so \$100 from either action & Joint earnings & States its own equilibrium mix correctly (5) \\
 & Column & Right & 5 & Predicts that Row is drawn to the large prize & Chooses Right as the best response to Top and cites the equilibrium weight & States its own equilibrium mix correctly (5) \\
\addlinespace
Gemini 3.8 Flash & Row & Bottom & 2 & Predicts that Column anticipates Row's pull toward the large prize & Chooses Bottom as the best response to Right & Names Goeree and Holt (2001) (1) \\
 & Row & Top & 8 & Computes the threshold $1/k$ for Left & Judges that the probability of Left exceeds $1/k$ & Names Goeree and Holt (2001) (1) \\
 & Column & Right & 5 & Predicts that Row is drawn to the large prize & Chooses Right as the best response to Top &  \\
 & Column & Right & 5 & Predicts that Row is drawn to the large prize & Chooses Right as the best response to Top and cites the equilibrium weight &  \\
\addlinespace
Claude Haiku 5.5 & Column & Left & 10 & Computes the equilibrium mix & Joint earnings & States its own equilibrium mix correctly (7)\newline States its own equilibrium mix as 1/2, incorrectly (2)\newline Leaves its own equilibrium mix incomplete (1) \\
 & Column & Right & 9 & Computes the equilibrium mix & Action with the larger equilibrium weight & States its own equilibrium mix correctly (9) \\
 & Column & Right & 1 & Computes the equilibrium mix & Larger equilibrium weight and stable joint earnings & States its own equilibrium mix correctly (1) \\
\end{longtable}

\clearpage
\section{Reply Validity and Truncation}\label{app:validity}

Baseline runs permit up to 256 output tokens for immediate answers and 16,384 with reasoning, except Qwen3.8's reasoning runs, which permit 32,768.

The open-weight runs also include a presentation control at $T=0$, with reasoning off and on. Greedy decoding supplies no sampling randomness for a fixed prompt, so these choices measure variation across presentations rather than within-prompt mixing. I schedule one reply per variant and cell: 12 per game cell, or 216 across the 18 cells in each response mode, and 12 per prescribed probability in the stated-strategy task, or 108 across its nine probabilities. These counts are scheduled draws rather than quotas of valid replies. The zero-temperature runs are reported only in this appendix and are excluded from the main paper's mixing tests, payoff-response estimates, choice shares, and figures.

Tables~\ref{tab:validity_game} and~\ref{tab:validity_task} classify every reported baseline draw by model, response mode, and temperature under the validity rules of Section~\ref{sec:subjects} of the paper. They give the band check that determines which temperatures are retained and report how often replies are truncated at the output cap. With immediate answers, truncation is concentrated at the highest temperatures and reaches 199 of 3,024 draws for Gemma~4 31B at $T=2$ in the game, a temperature that fails the band check. With reasoning on, the chat template of Qwen3.8 27B adds a default system message that sets its reasoning effort to ``xhigh,'' and I keep this default. Its replies are long, and at the cap of 16,384 tokens used for the other models it truncates 285 of the 3,024 game replies in its band check at $T=1$, nearly all at $k=4$ and $k=8$, so that temperature fails the check. Its reasoning runs therefore use a cap of 32,768 tokens, at which it truncates 7 of its 8,648 game replies and none of its 4,320 task replies at $T=1$. Nemotron~3.5 Lightning runs at the cap of 16,384 tokens and truncates 37 of its 8,678 game replies and none of its 4,320 task replies at $T=1$. Every recorded frontier reply in these baseline runs is valid. Validity for the randomizer-access treatment is reported separately in Appendix~\ref{app:random_tool}.

\footnotesize
\setlength{\tabcolsep}{3.5pt}
\begin{longtable}{@{}>{\raggedright\arraybackslash}p{0.14\textwidth}lrlrrrrrrr@{}}
\caption{Reply Validity in the Game}\label{tab:validity_game} \\
\toprule
& & & & Band & & Valid & Off- & Not & & \\
Model & Mode & $T$ & Status & check & Draws & (\%) & format & offered & Refusal & Truncated \\
\midrule
\endfirsthead
\multicolumn{11}{l}{\textit{Table \thetable, continued}} \\
\toprule
& & & & Band & & Valid & Off- & Not & & \\
Model & Mode & $T$ & Status & check & Draws & (\%) & format & offered & Refusal & Truncated \\
\midrule
\endhead
\bottomrule
\multicolumn{11}{@{}p{\textwidth}@{}}{\rule{0pt}{1.2em}\footnotesize \textit{Notes.} Draws are all recorded baseline replies, including the band-check block and redraws. Development checks and the randomizer comparison are reported separately. ``Band check'' is the percentage of valid replies in the first block of 14 draws per variant in every cell. A temperature is retained if this percentage is at least 95, and its quota is then filled. At a dropped temperature only the band-check block is drawn, and higher temperatures are skipped and do not appear. ``Greedy'' marks $T=0$, with one draw per variant and cell. ``API'' marks the frontier runs, which have no band check. Invalid replies are off-format, name a letter that was not offered, refuse, or are truncated at the output cap of 256 tokens with reasoning off or 16,384 tokens with reasoning on (32,768 for Qwen3.8 27B).} \\
\endlastfoot
Qwen3.8 27B & Immediate & 0 & Greedy &  & 216 & 100.0 & 0 & 0 & 0 & 0 \\
&  & 0.25 & Retained & 100.0 & 8,640 & 100.0 & 0 & 0 & 0 & 0 \\
&  & 0.5 & Retained & 100.0 & 8,640 & 100.0 & 0 & 0 & 0 & 0 \\
&  & 0.75 & Retained & 99.9 & 8,643 & 100.0 & 1 & 2 & 0 & 0 \\
&  & 1 & Retained & 99.6 & 8,679 & 99.6 & 8 & 15 & 0 & 16 \\
&  & 1.25 & Retained & 97.0 & 8,904 & 97.0 & 163 & 58 & 0 & 43 \\
&  & 1.5 & Dropped & 85.4 & 3,024 & 85.4 & 272 & 40 & 0 & 129 \\
\addlinespace
Gemma 4 31B & Immediate & 0 & Greedy &  & 216 & 100.0 & 0 & 0 & 0 & 0 \\
&  & 0.25 & Retained & 100.0 & 8,640 & 100.0 & 0 & 0 & 0 & 0 \\
&  & 0.5 & Retained & 100.0 & 8,640 & 100.0 & 0 & 0 & 0 & 0 \\
&  & 0.75 & Retained & 100.0 & 8,640 & 100.0 & 0 & 0 & 0 & 0 \\
&  & 1 & Retained & 100.0 & 8,640 & 100.0 & 0 & 0 & 0 & 0 \\
&  & 1.25 & Retained & 100.0 & 8,640 & 100.0 & 0 & 0 & 0 & 0 \\
&  & 1.5 & Retained & 99.7 & 8,657 & 99.8 & 15 & 0 & 0 & 2 \\
&  & 2 & Dropped & 87.0 & 3,024 & 87.0 & 192 & 0 & 2 & 199 \\
\addlinespace
Nemotron 3.5 Lightning & Immediate & 0 & Greedy &  & 216 & 100.0 & 0 & 0 & 0 & 0 \\
&  & 0.25 & Retained & 100.0 & 8,640 & 100.0 & 0 & 0 & 0 & 0 \\
&  & 0.5 & Retained & 100.0 & 8,640 & 100.0 & 0 & 0 & 0 & 0 \\
&  & 0.75 & Retained & 99.8 & 8,659 & 99.8 & 18 & 0 & 1 & 0 \\
&  & 1 & Retained & 98.2 & 8,808 & 98.1 & 139 & 1 & 7 & 21 \\
&  & 1.25 & Dropped & 89.7 & 3,024 & 89.7 & 194 & 5 & 3 & 108 \\
\addlinespace
Qwen3.8 27B & Reasoning & 0 & Greedy &  & 216 & 100.0 & 0 & 0 & 0 & 0 \\
&  & 1 & Retained & 99.8 & 8,648 & 99.9 & 0 & 1 & 0 & 7 \\
\addlinespace
Gemma 4 31B & Reasoning & 0 & Greedy &  & 216 & 99.5 & 0 & 0 & 0 & 1 \\
&  & 1 & Retained & 100.0 & 8,641 & 100.0 & 1 & 0 & 0 & 0 \\
\addlinespace
Nemotron 3.5 Lightning & Reasoning & 0 & Greedy &  & 216 & 99.1 & 0 & 0 & 0 & 2 \\
&  & 1 & Retained & 99.6 & 8,678 & 99.6 & 1 & 0 & 0 & 37 \\
\addlinespace
Gemini 3.8 Flash & Reasoning & 1 & API &  & 2,160 & 100.0 & 0 & 0 & 0 & 0 \\
\addlinespace
Claude Haiku 5.5 & Reasoning & Default & API &  & 2,160 & 100.0 & 0 & 0 & 0 & 0 \\
\addlinespace
GPT-6 Luna & Reasoning & 1 & API &  & 2,160 & 100.0 & 0 & 0 & 0 & 0 \\
\end{longtable}

\footnotesize
\setlength{\tabcolsep}{3.5pt}
\begin{longtable}{@{}>{\raggedright\arraybackslash}p{0.14\textwidth}lrlrrrrrrr@{}}
\caption{Reply Validity in the Stated-Strategy Task}\label{tab:validity_task} \\
\toprule
& & & & Band & & Valid & Off- & Not & & \\
Model & Mode & $T$ & Status & check & Draws & (\%) & format & offered & Refusal & Truncated \\
\midrule
\endfirsthead
\multicolumn{11}{l}{\textit{Table \thetable, continued}} \\
\toprule
& & & & Band & & Valid & Off- & Not & & \\
Model & Mode & $T$ & Status & check & Draws & (\%) & format & offered & Refusal & Truncated \\
\midrule
\endhead
\bottomrule
\multicolumn{11}{@{}p{\textwidth}@{}}{\rule{0pt}{1.2em}\footnotesize \textit{Notes.} Draws are all recorded baseline replies, including the band-check block and redraws. Development checks and the randomizer comparison are reported separately. ``Band check'' is the percentage of valid replies in the first block of 14 draws per variant in every cell. A temperature is retained if this percentage is at least 95, and its quota is then filled. At a dropped temperature only the band-check block is drawn, and higher temperatures are skipped and do not appear. ``Greedy'' marks $T=0$, with one draw per variant and cell. ``API'' marks the frontier runs, which have no band check. Invalid replies are off-format, name a letter that was not offered, refuse, or are truncated at the output cap of 256 tokens with reasoning off or 16,384 tokens with reasoning on (32,768 for Qwen3.8 27B).} \\
\endlastfoot
Qwen3.8 27B & Immediate & 0 & Greedy &  & 108 & 100.0 & 0 & 0 & 0 & 0 \\
&  & 0.25 & Retained & 100.0 & 4,320 & 100.0 & 0 & 0 & 0 & 0 \\
&  & 0.5 & Retained & 100.0 & 4,320 & 100.0 & 0 & 0 & 0 & 0 \\
&  & 0.75 & Retained & 100.0 & 4,320 & 100.0 & 0 & 0 & 0 & 0 \\
&  & 1 & Retained & 99.8 & 4,327 & 99.8 & 5 & 1 & 0 & 1 \\
&  & 1.25 & Retained & 98.2 & 4,402 & 98.1 & 65 & 10 & 0 & 7 \\
&  & 1.5 & Dropped & 85.5 & 1,512 & 85.5 & 144 & 13 & 0 & 62 \\
\addlinespace
Gemma 4 31B & Immediate & 0 & Greedy &  & 108 & 100.0 & 0 & 0 & 0 & 0 \\
&  & 0.25 & Retained & 100.0 & 4,320 & 100.0 & 0 & 0 & 0 & 0 \\
&  & 0.5 & Retained & 100.0 & 4,320 & 100.0 & 0 & 0 & 0 & 0 \\
&  & 0.75 & Retained & 100.0 & 4,320 & 100.0 & 0 & 0 & 0 & 0 \\
&  & 1 & Retained & 100.0 & 4,320 & 100.0 & 0 & 0 & 0 & 0 \\
&  & 1.25 & Retained & 100.0 & 4,322 & 100.0 & 2 & 0 & 0 & 0 \\
&  & 1.5 & Retained & 99.9 & 4,331 & 99.7 & 11 & 0 & 0 & 0 \\
&  & 2 & Dropped & 80.0 & 1,512 & 80.0 & 199 & 0 & 1 & 102 \\
\addlinespace
Nemotron 3.5 Lightning & Immediate & 0 & Greedy &  & 108 & 100.0 & 0 & 0 & 0 & 0 \\
&  & 0.25 & Retained & 100.0 & 4,320 & 100.0 & 0 & 0 & 0 & 0 \\
&  & 0.5 & Retained & 99.9 & 4,321 & 100.0 & 1 & 0 & 0 & 0 \\
&  & 0.75 & Retained & 99.3 & 4,344 & 99.4 & 17 & 2 & 2 & 3 \\
&  & 1 & Retained & 97.1 & 4,455 & 97.0 & 92 & 7 & 24 & 12 \\
&  & 1.25 & Dropped & 82.9 & 1,512 & 82.9 & 166 & 2 & 19 & 72 \\
\addlinespace
Qwen3.8 27B & Reasoning & 0 & Greedy &  & 108 & 100.0 & 0 & 0 & 0 & 0 \\
&  & 1 & Retained & 100.0 & 4,320 & 100.0 & 0 & 0 & 0 & 0 \\
\addlinespace
Gemma 4 31B & Reasoning & 0 & Greedy &  & 108 & 100.0 & 0 & 0 & 0 & 0 \\
&  & 1 & Retained & 100.0 & 4,320 & 100.0 & 0 & 0 & 0 & 0 \\
\addlinespace
Nemotron 3.5 Lightning & Reasoning & 0 & Greedy &  & 108 & 100.0 & 0 & 0 & 0 & 0 \\
&  & 1 & Retained & 100.0 & 4,320 & 100.0 & 0 & 0 & 0 & 0 \\
\addlinespace
Claude Haiku 5.5 & Reasoning & Default & API &  & 1,080 & 100.0 & 0 & 0 & 0 & 0 \\
\addlinespace
GPT-6 Luna & Reasoning & 1 & API &  & 1,080 & 100.0 & 0 & 0 & 0 & 0 \\
\addlinespace
Gemini 3.8 Flash & Reasoning & 1 & API &  & 1,080 & 100.0 & 0 & 0 & 0 & 0 \\
\end{longtable}

\clearpage
\section{Randomizer-Access Treatment}\label{app:random_tool}

\subsection{Protocol}

Section~\ref{sec:results} of the paper compares each model's reasoning replies with and without access to a randomizer. The access arm has the design of the reasoning baseline, with 10 valid decisions per presentation for the frontier models and 40 for the open-weight models in each of the 18 game cells and at each of the nine prescribed probabilities, or 3,240 and 12,960 decisions per model. The reasoning baseline is the control. Every decision begins a new conversation, and every inference setting copies the model's baseline run, namely the cap of 16,384 output tokens per turn (32,768 for Qwen3.8 27B, and for Nemotron~3.5 Lightning as described below), the provider's default reasoning level for Claude Haiku~5.5 and Gemini~3.8 Flash, medium effort for GPT-6 Luna, and $T=1$ wherever the API accepts a temperature. Invalid decisions are redrawn up to 1.25 times the quota, as in the baseline. Haiku and Luna each return one reply without the closing sentence, which is redrawn, and every other reply is valid. Gemma~4 runs locally in bf16 at $T=1$ with a top-$p$ of one, the same cap, and a context of 19,456 tokens, and all 12,960 of its final replies are valid. Qwen3.8 runs the same way with its larger cap and a context of 65,536 tokens. Of its 13,278 final replies, 12,960 are valid, 301 are truncated, 15 are off-format, and 2 name a letter that was not offered. Nemotron first ran with its baseline cap of 16,384 tokens, and only 85.5 percent of its first 3,024 decisions were valid, below the 95 percent rule of the baseline. Most invalid replies were truncated or ended with a draw request set in Markdown. I therefore restarted its run with the cap of 32,768 tokens, a context of 49,152 tokens, and the request rule described below, which I adopted after this first run. Of its 13,283 final replies, 12,960 are valid, 309 are off-format, 7 name a letter that was not offered, 4 decline to choose, 2 are truncated, and 1 reaches the turn limit.

The control uses the original reasoning prompt. In the access arm, the following paragraph is inserted between the unchanged task body and its unchanged closing instruction:
\begin{quote}\small
A random\_uniform tool is available if you find it useful. It returns a fresh random number between 0 (inclusive) and 1 (exclusive). You may request up to 4 draws for this decision. You decide whether to use it, how to use any returned numbers, and which letter to choose. Your final answer, rather than a tool result, is your decision.
\end{quote}

The native function \texttt{random\_uniform()} takes no arguments and returns only a number. It supplies no distribution, threshold, action, or equilibrium advice. The model chooses whether to call it, how to map a draw to a letter, and whether to follow that mapping. The game prompt does not provide the equilibrium probabilities or require randomization. Only the last non-tool visible reply is parsed for the final letter. Intermediate text, reasoning summaries, and tool results are not actions. The limit is four tool requests and six model turns per decision, and every requested draw is retained. The native interfaces are Anthropic tool use, OpenAI Responses function calling, and Google generateContent function calling. Only Qwen3.8 and Nemotron ever use a second draw, in 30 and 17 of their valid decisions, and only one of Nemotron's valid decisions uses a third.

The open-weight models run locally and request draws in text rather than through a native function-calling interface, so their access paragraph ends with one further instruction, ``To request a draw, end your current message with exactly CALL random\_uniform() on its own final line. A tool-result message will supply the number, and you can then continue your response. Do not give your final answer until you have finished any tool requests.'' The host treats a message as a request when its final line holds only this text, plain or set in Markdown bold, italics, code, or a heading. This rule reclassifies no reply of Gemma~4 or Qwen3.8, which always write the line exactly as instructed, and it covers 868 of Nemotron's 12,541 requests. The host returns the draw as a new message, and the model continues the same conversation. The chat templates of Gemma~4 and Nemotron remove reasoning from earlier turns, so after a draw each sees its earlier visible text and the number but not its earlier reasoning. The template of Qwen3.8 keeps earlier reasoning by default, as the frontier interfaces do, and I keep this default. Every request receives a draw, except two requests in one Nemotron decision that exceed the limit of four.

The host generates reproducible pseudorandom uniform draws with seeds derived from the frozen master seed, decision identifier, and draw ordinal. The master seeds are 2026101701 for Haiku, 2026101703 for Luna, 2026101705 for Gemini, 2026101706 for Gemma~4, 2026101707 for Qwen3.8, and 2026101708 for Nemotron, and seeds are never shown to the model. Frozen configurations, rendered prompts, original provider responses, draw values, and source snapshots are archived under \texttt{data/\allowbreak raw/\allowbreak random\_tool/} in the replication materials. The interaction audit replays every draw, checks the normalized replies against the original provider responses, and verifies that each recorded action is the final model reply.

\subsection{Choice Frequencies and Tool Use}

Table~\ref{tab:random_tool_panel} reports final choices with and without access. For each task, the mean absolute frequency error is the equally weighted mean of $|\hat f_c-p_c|$ across its cells, where $\hat f_c$ is the observed share and $p_c$ is the stated probability or equilibrium target. These errors include sampling variation, so the table also reports the error that exact implementation of every target would show at the same sample sizes, about 2.5 percentage points in the stated-strategy task and 1.9 in the game for the frontier models, and 1.3 and 1.0 for the open-weight models. With access, the stated-strategy errors of all six models and the game error of Haiku lie near or below this benchmark, while the game errors of Luna, Nemotron, Qwen3.8, Gemini, and Gemma~4 lie above it.

Haiku draws in 2,139 of its 2,160 game decisions. Relative to its reasoning replies without access, its Row slope in equation~(1) of the paper falls from 0.33 to 0.02 (wild bootstrap $p=0.03$ for the change), and its mean absolute frequency error across the six game cells falls from 17.31 to 2.22 percentage points, near the 1.9 points that exact equilibrium play would show at this sample size. Luna draws in 85 percent of its decisions at $k=4$ and $k=8$, its Row slope falls from 0.35 to 0.04 ($p=0.02$), and its error falls from 19.49 to 4.81 points.

Nemotron draws in 94 percent of its game decisions at $k=4$ and $k=8$, and its Row slope falls from 0.34 to 0.02 (wild bootstrap $p=0.02$ for the change). As Column, however, it chooses Left in 33.0 and 30.6 percent of replies at these values, so its mean absolute frequency error across the six game cells falls from 25.19 to 5.43 percentage points, against 1.0 point under exact equilibrium play at its larger sample.

Qwen3.8 draws in 96 percent of its game decisions at $k=2$ but in 62 percent at $k=4$ and $k=8$. Its Row slope falls from 0.35 to 0.15 (wild bootstrap $p=0.07$ for the change), and its mean absolute frequency error across the six game cells falls from 22.23 to 8.46 percentage points.

Gemini draws in all 720 of its game decisions at $k=2$ but in 271 of 720 at $k=8$. Its Row slope falls from 0.36 to 0.19 (wild bootstrap $p=0.15$ for the change), it chooses Left as Column in 14 of 720 replies at $k=4$ and $k=8$, and its mean absolute frequency error across the six game cells falls from 17.08 to 12.04 percentage points.

Gemma~4 draws in 2,829 of its 2,880 game decisions at $k=2$ but in only 393 of 5,760 at $k=4$ and $k=8$. Its Row slope changes from 0.36 to 0.35 (wild bootstrap $p=0.89$ for the change), and its mean absolute frequency error across the six game cells rises from 18.45 to 19.14 percentage points.

Table~\ref{tab:random_tool_panel_use} reports tool use. Four of the six models draw in every stated-strategy decision, Qwen3.8 draws in 4,306 of 4,320, and Nemotron in 4,300. The final action follows the rule ``choose the named action if the draw is below $\pi$'' in 1,066 to 1,076 of 1,080 decisions for the frontier models, in 4,088 of 4,320 for Gemma~4, in 4,232 of 4,320 for Qwen3.8, and in 4,229 of 4,320 for Nemotron. Every other decision of the first four models follows the reverse rule ``choose the named action if the draw is at least $1-\pi$,'' which implements the same probability, as do 66 of Qwen3.8's and the other 71 of Nemotron's with a draw, while 8 of Qwen3.8's decisions with a draw follow neither rule. Gemma~4 uses the reverse rule in 201 of its 480 decisions at $\pi=1/2$, where either rule fits, and in 31 of 3,840 elsewhere. In the game, Haiku draws in nearly every decision, while Gemini draws in 136 of 360 Row decisions and 135 of 360 Column decisions at $k=8$, Qwen3.8 in 790 and 835 of 1,440, and Nemotron in 1,329 and 1,348 of 1,440. Gemma~4 draws in 2,829 of its 2,880 decisions at $k=2$ but in 89 and 21 of 1,440 Row decisions and 240 and 43 of 1,440 Column decisions at $k=4$ and $k=8$. Since calling the tool is the model's choice, caller--noncaller contrasts do not identify an effect of calling. The access contrast also includes the paragraph describing the tool and the opportunity to continue after receiving a draw, so it does not separately identify the effect of the numeric draw.

\begin{table}[p]
\centering
\footnotesize
\setlength{\tabcolsep}{2pt}
\caption{Final Choices by Randomizer Access, Panel Runs}\label{tab:random_tool_panel}
\begin{tabular*}{\textwidth}{@{\extracolsep{\fill}}lrrrrrrrrrrrrr@{}}
\toprule
& & \multicolumn{2}{c}{Haiku} & \multicolumn{2}{c}{Luna} & \multicolumn{2}{c}{Gemini} & \multicolumn{2}{c}{Gemma 4} & \multicolumn{2}{c}{Qwen3.8} & \multicolumn{2}{c}{Nemotron} \\
\cmidrule(lr){3-4}\cmidrule(lr){5-6}\cmidrule(lr){7-8}\cmidrule(lr){9-10}\cmidrule(lr){11-12}\cmidrule(lr){13-14}
Cell & Target & None & Avail. & None & Avail. & None & Avail. & None & Avail. & None & Avail. & None & Avail. \\
\midrule
\multicolumn{14}{l}{\textit{A. Stated probabilities}} \\
$\pi=1/100$ & 1.0 & 0.0 & 3.3 & 0.0 & 0.8 & 0.0 & 0.8 & 0.0 & 1.0 & 0.0 & 1.0 & 0.2 & 0.8 \\
$\pi=1/10$ & 10.0 & 0.0 & 10.0 & 0.0 & 11.7 & 0.0 & 7.5 & 0.0 & 10.2 & 0.0 & 9.0 & 2.1 & 10.6 \\
$\pi=1/4$ & 25.0 & 0.0 & 27.5 & 1.7 & 27.5 & 0.0 & 16.7 & 0.0 & 23.5 & 0.0 & 23.3 & 1.9 & 25.8 \\
$\pi=1/3$ & 33.3 & 0.0 & 40.0 & 3.3 & 36.7 & 0.0 & 37.5 & 0.0 & 31.0 & 0.0 & 37.3 & 3.8 & 31.9 \\
$\pi=1/2$ & 50.0 & 62.5 & 50.0 & 84.2 & 42.5 & 50.8 & 53.3 & 70.0 & 48.5 & 76.7 & 50.4 & 89.6 & 47.1 \\
$\pi=2/3$ & 66.7 & 100.0 & 65.0 & 100.0 & 70.0 & 100.0 & 65.8 & 100.0 & 66.0 & 100.0 & 66.2 & 98.8 & 67.3 \\
$\pi=3/4$ & 75.0 & 100.0 & 73.3 & 100.0 & 71.7 & 100.0 & 68.3 & 100.0 & 74.2 & 100.0 & 74.0 & 100.0 & 75.8 \\
$\pi=9/10$ & 90.0 & 100.0 & 93.3 & 100.0 & 90.0 & 100.0 & 89.2 & 100.0 & 88.7 & 100.0 & 93.1 & 100.0 & 88.3 \\
$\pi=99/100$ & 99.0 & 100.0 & 100.0 & 100.0 & 96.7 & 100.0 & 99.2 & 100.0 & 98.8 & 100.0 & 98.3 & 99.6 & 98.8 \\
\addlinespace
Mean absolute error & -- & 16.80 & 2.13 & 18.65 & 2.69 & 15.50 & 3.00 & 17.63 & 0.94 & 18.37 & 1.37 & 18.74 & 1.04 \\
Exact implementation & -- & 2.51 & 2.51 & 2.51 & 2.50 & 2.51 & 2.51 & 1.25 & 1.26 & 1.26 & 1.26 & 1.26 & 1.26 \\
\midrule
\multicolumn{14}{l}{\textit{B. Asymmetric matching pennies}} \\
Row, $k=2$ & 50.0 & 49.7 & 51.9 & 49.7 & 52.2 & 48.9 & 51.7 & 50.0 & 51.0 & 50.3 & 52.8 & 50.1 & 50.1 \\
Row, $k=4$ & 50.0 & 98.6 & 49.7 & 97.8 & 57.8 & 63.9 & 41.7 & 99.8 & 96.3 & 99.4 & 64.8 & 97.2 & 52.6 \\
Row, $k=8$ & 50.0 & 95.0 & 55.0 & 97.5 & 58.3 & 98.9 & 78.1 & 99.7 & 99.0 & 99.4 & 73.7 & 96.9 & 52.8 \\
Column, $k=2$ & 50.0 & 49.7 & 50.8 & 47.8 & 50.0 & 51.1 & 49.4 & 50.2 & 50.0 & 49.3 & 49.8 & 50.1 & 49.1 \\
Column, $k=4$ & 25.0 & 19.4 & 28.3 & 33.6 & 32.5 & 0.0 & 3.6 & 16.2 & 9.9 & 2.3 & 19.7 & 47.4 & 33.0 \\
Column, $k=8$ & 12.5 & 16.7 & 10.6 & 23.1 & 15.6 & 0.0 & 0.3 & 14.7 & 9.0 & 1.6 & 8.5 & 46.9 & 30.6 \\
\addlinespace
Mean absolute error & -- & 17.31 & 2.22 & 19.49 & 4.81 & 17.08 & 12.04 & 18.45 & 19.14 & 22.23 & 8.46 & 25.19 & 5.43 \\
Exact implementation & -- & 1.94 & 1.94 & 1.94 & 1.93 & 1.94 & 1.94 & 0.97 & 0.97 & 0.97 & 0.97 & 0.97 & 0.97 \\
\addlinespace
\bottomrule
\end{tabular*}
\vspace{0.5em}
\begin{minipage}{\textwidth}\footnotesize
\textit{Notes.} Entries are percentages of valid final choices. Each cell has 120 choices in Panel A and 360 in Panel B for the frontier models, and 480 and 1{,}440 for the open-weight models. Panel B pools the base prizes \$10, \$100, and \$1{,}000. ``None'' is the model's reasoning baseline without access, and ``Avail.'' the panel run with access to the number-only randomizer, with the same settings. The named action is measured in Panel A, Top for Row and Left for Column in Panel B. Shares weight the 12 presentations equally. Mean absolute errors weight each panel's cells equally and are in percentage points. ``Exact implementation'' is the mean error that a player choosing each action with exactly the target probability would show at the same sample sizes, from 200{,}000 simulated samples.
\end{minipage}
\end{table}

\begin{table}[p]
\centering
\footnotesize
\setlength{\tabcolsep}{2pt}
\caption{Randomizer Use, Panel Runs}\label{tab:random_tool_panel_use}
\begin{tabular*}{\textwidth}{@{\extracolsep{\fill}}lrrrrrr@{}}
\toprule
Cell & Haiku & Luna & Gemini & Gemma 4 & Qwen3.8 & Nemotron \\
\midrule
\multicolumn{7}{l}{\textit{A. Asymmetric matching pennies}} \\
Decisions per cell & 360 & 360 & 360 & 1,440 & 1,440 & 1,440 \\
Row, $k=2$ & 359 & 358 & 360 & 1,419 & 1,384 & 1,394 \\
Row, $k=4$ & 357 & 276 & 337 & 89 & 950 & 1,358 \\
Row, $k=8$ & 351 & 289 & 136 & 21 & 790 & 1,329 \\
Column, $k=2$ & 353 & 356 & 360 & 1,410 & 1,387 & 1,408 \\
Column, $k=4$ & 360 & 344 & 287 & 240 & 1,018 & 1,376 \\
Column, $k=8$ & 359 & 322 & 135 & 43 & 835 & 1,348 \\
\addlinespace
\multicolumn{7}{l}{\textit{B. Stated probabilities}} \\
Decisions & 1,080 & 1,080 & 1,080 & 4,320 & 4,320 & 4,320 \\
Decisions with a draw & 1,080 & 1,080 & 1,080 & 4,320 & 4,306 & 4,300 \\
Named action iff draw $<\pi$ & 1,076 & 1,075 & 1,066 & 4,088 & 4,232 & 4,229 \\
Named action iff draw $\geq 1-\pi$ only & 4 & 5 & 14 & 232 & 66 & 71 \\
Neither rule & 0 & 0 & 0 & 0 & 8 & 0 \\
\bottomrule
\end{tabular*}
\vspace{0.5em}
\begin{minipage}{\textwidth}\footnotesize
\textit{Notes.} Panel~A counts decisions in which the model requests and receives a draw, pooling the three base prizes. Every caller requests one draw, except in 30 Qwen3.8 and 17 Nemotron decisions with more, which are classified by the first. Panel~B also classifies each stated-strategy decision with a draw by the rule its final action fits. The two rules implement the same probability, and at $\pi=1/2$ every action fits one of them. Model names follow Table~\ref{tab:random_tool_panel}.
\end{minipage}
\end{table}

\clearpage
\section{Additional Game Results}\label{app:game_more}

This appendix reports game results that Section~\ref{sec:game_results} and the conclusion of the paper summarize. As in Figure~\ref{fig:game_results} of the paper, game shares pool the three base prizes at $T=1$.

\subsection{Row's Own-Payoff Response}

Using equation~(1) of the paper, the Row slope with reasoning is 0.35 (standard error 0.11, wild bootstrap $p=0.002$) for Qwen3.8, 0.36 (0.11, $p<0.001$) for Gemma~4, 0.34 (0.10, $p=0.006$) for Nemotron, 0.33 (0.10, $p=0.01$) for Haiku, 0.35 (0.11, $p=0.02$) for Luna, and 0.36 (0.10, $p<0.001$) for Gemini. These estimates pool the three base prizes and use variant fixed effects, with standard errors clustered by variant and wild bootstrap tests of the equilibrium prediction of a zero slope.

\subsection{Column with Reasoning}

With reasoning, Gemma~4 chooses Left in 16.2 and 14.7 percent of replies at $k=4$ and $k=8$, respectively, Haiku in 19.4 and 16.7 percent, Qwen3.8 in 2.3 and 1.6 percent, and Gemini in none at either value. These shares pool the three base prizes, as in Figure~\ref{fig:game_results} of the paper.

\subsection{Tests against Equilibrium}

Table~\ref{tab:equilibrium_tests} tests each share plotted in Figure~\ref{fig:game_results} at $k=4$ and $k=8$ against the equilibrium probability. At $k=2$ the test carries no information, since paired variants present the same prompt and the pooled share is one half by construction. With reasoning, the tests reject equilibrium in all 24 cells. With immediate answers, they reject in all 12 cells. With the randomizer, the tests reject in none of Haiku's four cells, in the two Column cells of Nemotron, in three of Luna's, and in all four of Gemini's, Gemma~4's, and Qwen3.8's. A test that does not reject does not show that play is at equilibrium.

\begin{table}[htbp]
\centering
\footnotesize
\setlength{\tabcolsep}{2.5pt}
\caption{Game Shares Tested against Equilibrium}\label{tab:equilibrium_tests}
\begin{tabular*}{\textwidth}{@{\extracolsep{\fill}}llrrrrrrrrr}
\toprule
& & & \multicolumn{2}{c}{Row, $k=4$} & \multicolumn{2}{c}{Row, $k=8$} & \multicolumn{2}{c}{Column, $k=4$} & \multicolumn{2}{c}{Column, $k=8$} \\
\cmidrule(lr){4-5}\cmidrule(lr){6-7}\cmidrule(lr){8-9}\cmidrule(lr){10-11}
Model & Mode & $n$ & Share & $p$ & Share & $p$ & Share & $p$ & Share & $p$ \\
\midrule
Qwen3.8 27B & Immediate & 1,440 & 65.2 & $<$0.001 & 65.3 & $<$0.001 & 56.3 & $<$0.001 & 52.4 & $<$0.001 \\
Qwen3.8 27B & Reasoning & 1,440 & 99.4 & $<$0.001 & 99.4 & $<$0.001 & 2.3 & $<$0.001 & 1.6 & $<$0.001 \\
Qwen3.8 27B & Randomizer & 1,440 & 64.8 & $<$0.001 & 73.7 & $<$0.001 & 19.7 & $<$0.001 & 8.5 & $<$0.001 \\
\addlinespace
Gemma 4 31B & Immediate & 1,440 & 55.8 & $<$0.001 & 97.2 & $<$0.001 & 32.5 & $<$0.001 & 28.5 & $<$0.001 \\
Gemma 4 31B & Reasoning & 1,440 & 99.8 & $<$0.001 & 99.7 & $<$0.001 & 16.2 & $<$0.001 & 14.7 & 0.012 \\
Gemma 4 31B & Randomizer & 1,440 & 96.3 & $<$0.001 & 99.0 & $<$0.001 & 9.9 & $<$0.001 & 9.0 & $<$0.001 \\
\addlinespace
Nemotron 3.5 Lightning & Immediate & 1,440 & 54.2 & 0.001 & 55.4 & $<$0.001 & 55.1 & $<$0.001 & 53.7 & $<$0.001 \\
Nemotron 3.5 Lightning & Reasoning & 1,440 & 97.2 & $<$0.001 & 96.9 & $<$0.001 & 47.4 & $<$0.001 & 46.9 & $<$0.001 \\
Nemotron 3.5 Lightning & Randomizer & 1,440 & 52.6 & 0.066 & 52.8 & 0.066 & 33.0 & $<$0.001 & 30.6 & $<$0.001 \\
\addlinespace
Claude Haiku 5.5 & Reasoning & 360 & 98.6 & $<$0.001 & 95.0 & $<$0.001 & 19.4 & 0.030 & 16.7 & 0.030 \\
Claude Haiku 5.5 & Randomizer & 360 & 49.7 & 0.958 & 55.0 & 0.260 & 28.3 & 0.433 & 10.6 & 0.600 \\
\addlinespace
GPT-6 Luna & Reasoning & 360 & 97.8 & $<$0.001 & 97.5 & $<$0.001 & 33.6 & $<$0.001 & 23.1 & $<$0.001 \\
GPT-6 Luna & Randomizer & 360 & 57.8 & 0.007 & 58.3 & 0.006 & 32.5 & 0.006 & 15.6 & 0.094 \\
\addlinespace
Gemini 3.8 Flash & Reasoning & 360 & 63.9 & $<$0.001 & 98.9 & $<$0.001 & 0.0 & $<$0.001 & 0.0 & $<$0.001 \\
Gemini 3.8 Flash & Randomizer & 360 & 41.7 & 0.002 & 78.1 & $<$0.001 & 3.6 & $<$0.001 & 0.3 & $<$0.001 \\
\bottomrule
\end{tabular*}
\begin{minipage}{\textwidth}
\vspace{0.4em}\footnotesize \textit{Notes.} Shares are percentages choosing Top (Row) or Left (Column), pooled over the three base prizes, with $n$ valid replies per cell. The equilibrium is 50 percent for Row and 25 and 12.5 percent for Column. $p$ is the Holm-adjusted exact binomial $p$-value within each row.
\end{minipage}
\end{table}

\subsection{Prompt Variants}

With reasoning, play varies little across the 12 variants once the payoffs favor one action. At $k=4$ and $k=8$, chi-square tests reject a common probability across the variants in only 3 of the 42 cells of Gemma~4 as Column and the three frontier models in either role, about what chance produces at the 5 percent level. Within prompts, play is often pure. With reasoning, 65 of Gemma~4's 72 Row prompt versions at $k=4$ and $k=8$, 12 variants at each of three base prizes, choose Top in all 40 replies.

\subsection{Column with Immediate Answers}

With immediate answers, Column play responds little to Row's prize. Qwen3.8 chooses Left in 56.3 and 52.4 percent of replies at $k=4$ and $k=8$, and Nemotron in 55.1 and 53.7 percent, which fits a Column player who treats Row as unresponsive to Row's prizes, as human subjects often do \citep{weizsacker2003ignoring}. Gemma~4 moves partway, to 32.5 and 28.5 percent, with a slope from equation~(1) of the paper of $-0.16$ (standard error 0.13, wild bootstrap $p=0.28$). Its Row slope with immediate answers is 0.34 (0.10, $p<0.001$), against 0.12 (0.10, $p=0.22$) for Qwen3.8 and 0.03 (0.03, $p=0.29$) for Nemotron.

\subsection{Base Prize}

Holding $k$ and the variant fixed, a tenfold increase in the base prize moves a model's average share by at most 4.4 percentage points in either role, at any retained temperature, as the scale-free equilibrium predicts. Payoff scale likewise has little systematic effect on human play in similar games \citep{mckelvey2000effects}. The particular amount still shifts some cells without a trend. Gemma~4's immediate answers as Column at $k=8$ choose Left in 20.8, 39.8, and 24.8 percent of replies at $u=\$10$, \$100, and \$1,000, and the \$100 cell stands out at every retained temperature. Luna's reasoning replies in the same cell choose Left in 17.5, 32.5, and 19.2 percent. The amount thus acts on choices as presentation does rather than as an incentive.

\subsection{Temperature}

Only the open-weight models answering immediately vary temperature. Table~\ref{tab:temperature} shows that each prompt's own choices become less deterministic as temperature rises. From $T=0.25$ to the highest retained temperature, the average share of a prompt version's replies that choose its less common action rises from 9.3 to 31.4 percent for Qwen3.8, from 18.4 to 36.1 percent for Nemotron, and from 1.5 to 6.9 percent for Gemma~4, and the number of pure game versions falls from 122 to 2 of 216, from 75 to 0, and from 201 to 152. The added variation moves choices toward one half whatever the equilibrium. Gemma~4's share of Left as Column at $k=8$ rises from 28.1 to 31.7 percent, away from the equilibrium of 12.5 percent, and the shares of Qwen3.8 and Nemotron stay above one half. In the stated-strategy task, Qwen3.8's mean absolute error falls from 17.6 to 9.5 points as its share of the less likely action rises from 1.6 to 19.8 percent, against 17.3 under exact implementation, while Gemma~4's error moves by less than one point and it never chooses the less likely action. Nemotron's error falls from 36.7 to 29.8 points, but it chooses the less likely action in about 37 percent of replies at every temperature, because it favors the $\pi$-named action whatever its probability.

For the comparison reported in Section~\ref{sec:diagnostic} of the paper, increasing temperature from $T=0.25$ to $T=1$ lowers Qwen3.8's mean absolute error in the stated-strategy task from 17.6 to 9.1 percentage points. This comparison uses immediate answers.

\begin{table}[htbp]
\centering
\footnotesize
\setlength{\tabcolsep}{2.5pt}
\caption{Immediate Answers by Temperature}\label{tab:temperature}
\begin{tabular*}{\textwidth}{@{\extracolsep{\fill}}lrrrrrrr}
\toprule
& \multicolumn{5}{c}{Game} & \multicolumn{2}{c}{Stated-strategy task} \\
\cmidrule(lr){2-6}\cmidrule(lr){7-8}
$T$ & Less common & Pure of 216 & First listed & Top, $k=8$ & Left, $k=8$ & Error & Less likely \\
\midrule
\multicolumn{8}{l}{\textit{Qwen3.8 27B}} \\
0.25 & 9.3 & 122 & 64.6 & 76.7 & 57.2 & 17.6 & 1.6 \\
0.5 & 17.7 & 49 & 61.7 & 75.0 & 56.5 & 14.3 & 6.1 \\
0.75 & 23.6 & 20 & 61.3 & 70.0 & 54.2 & 10.7 & 11.4 \\
1 & 28.9 & 7 & 58.7 & 65.3 & 52.4 & 9.1 & 14.7 \\
1.25 & 31.4 & 2 & 57.4 & 65.1 & 54.2 & 9.5 & 19.8 \\
\addlinespace
\multicolumn{8}{l}{\textit{Gemma 4 31B}} \\
0.25 & 1.5 & 201 & 79.0 & 97.2 & 28.1 & 18.2 & 0.0 \\
0.5 & 2.4 & 196 & 78.9 & 97.4 & 26.7 & 18.5 & 0.0 \\
0.75 & 3.6 & 186 & 78.9 & 97.5 & 27.5 & 18.3 & 0.0 \\
1 & 4.5 & 176 & 78.4 & 97.2 & 28.5 & 18.2 & 0.0 \\
1.25 & 5.7 & 162 & 77.9 & 97.3 & 29.4 & 18.0 & 0.0 \\
1.5 & 6.9 & 152 & 77.7 & 96.5 & 31.7 & 18.0 & 0.0 \\
\addlinespace
\multicolumn{8}{l}{\textit{Nemotron 3.5 Lightning}} \\
0.25 & 18.4 & 75 & 49.4 & 60.0 & 59.2 & 36.7 & 36.6 \\
0.5 & 27.9 & 16 & 50.1 & 57.2 & 57.2 & 34.5 & 36.1 \\
0.75 & 33.4 & 0 & 50.8 & 56.7 & 56.8 & 33.2 & 37.6 \\
1 & 36.1 & 0 & 49.9 & 55.4 & 53.7 & 29.8 & 37.8 \\
\bottomrule
\end{tabular*}
\begin{minipage}{\textwidth}
\vspace{0.4em}\footnotesize \textit{Notes.} Percentages at each retained temperature. ``Less common'' is the mean share of a prompt version's replies that choose its less common action, and ``Pure'' counts versions with at least 95 percent on one action. Game shares pool the three base prizes. ``Error'' is the mean absolute error across the nine prescribed probabilities, and ``Less likely'' is the mean share choosing the less likely action at unequal probabilities, which is 17.3 under exact implementation.
\end{minipage}
\end{table}

\subsection{Exploitability}

Table~\ref{tab:exploitability} reports the gains available to an opponent fitted to each model's choices, with and without randomizer access. For each prompt, I select the opponent's action using half the replies and evaluate its payoff on the remaining replies. The table also gives confidence bounds for an oracle that knows each prompt's choice distribution.

\begin{table}[htbp]
\centering
\footnotesize
\setlength{\tabcolsep}{3pt}
\caption{Opponent Payoff Gains with and without Randomizer Access}\label{tab:exploitability}
\begin{tabular*}{\textwidth}{@{\extracolsep{\fill}}lrrr}
\toprule
Model & \makecell[r]{Without\\randomizer} & \makecell[r]{With\\randomizer} & \makecell[r]{Oracle bounds\\with randomizer} \\
\midrule
Qwen3.8 27B & 21.5 [20.2, 22.8] & 3.9 [2.1, 5.8] & [0.0, 9.5] \\
Gemma~4 31B & 24.1 [21.3, 26.9] & 11.1 [8.9, 13.3] & [6.3, 16.3] \\
Nemotron 3.5 Lightning & 56.2 [53.4, 59.0] & 13.7 [10.1, 17.3] & [8.7, 18.8] \\
Claude Haiku 5.5 & 25.8 [20.1, 31.4] & -1.5 [-7.3, 4.3] & [0.0, 14.3] \\
GPT-6 Luna & 23.6 [17.9, 29.3] & 3.7 [-1.9, 9.3] & [0.0, 18.5] \\
Gemini 3.8 Flash & 15.9 [13.3, 18.5] & 6.6 [4.0, 9.2] & [0.0, 15.5] \\
\bottomrule
\end{tabular*}
\begin{minipage}{\textwidth}
\vspace{0.4em}\footnotesize \textit{Notes.} Gains are percentages of the opponent's equilibrium payoff, averaged equally across both roles, the three prize multipliers, base prizes, and presentation variants. The 95 percent confidence intervals are conservative and pointwise across model and treatment estimates. Negative fitted-response gains are retained. Oracle bounds account for finite-sample response selection. These analyses were defined after reading replies.
\end{minipage}
\end{table}

\subsection{Agreement between Delegates}

Table~\ref{tab:agreement} reports how often two delegates of the same model, given the same prompt, choose the same action. Each reply is a separate request, so the replies to one prompt version stand for independent delegates that share a model. The rate is one half for Row in equilibrium and rises toward one at the equilibrium probabilities of Column as $k$ grows. With reasoning but no randomizer, Row agreement is at least 83 percent in every game for every model except Gemini at $k=4$. With the randomizer, it falls to about one half for Haiku, Luna, and Nemotron and at $k=2$ for Gemma~4, whose delegates still agree in 93 and 98 percent of Row decisions at $k=4$ and $k=8$. Gemini's delegates still agree in 66 percent of Row decisions at $k=8$ and in nearly all Column decisions at $k=4$ and $k=8$.

\begin{table}[htbp]
\centering
\footnotesize
\caption{Agreement between Delegates of the Same Model}\label{tab:agreement}
\begin{tabular*}{\textwidth}{@{\extracolsep{\fill}}llrrrrrr}
\toprule
& & \multicolumn{3}{c}{Row} & \multicolumn{3}{c}{Column} \\
\cmidrule(lr){3-5}\cmidrule(lr){6-8}
Model & Mode & $k=2$ & $k=4$ & $k=8$ & $k=2$ & $k=4$ & $k=8$ \\
\midrule
Qwen3.8 27B & Immediate & 77 & 60 & 60 & 64 & 54 & 55 \\
 & Reasoning & 99 & 99 & 99 & 58 & 95 & 97 \\
 & Randomizer & 50 & 55 & 61 & 50 & 68 & 84 \\
\addlinespace
Gemma~4 31B & Immediate & 100 & 97 & 98 & 100 & 83 & 84 \\
 & Reasoning & 100 & 100 & 99 & 84 & 73 & 75 \\
 & Randomizer & 50 & 93 & 98 & 50 & 82 & 84 \\
\addlinespace
Nemotron 3.5 Lightning & Immediate & 55 & 55 & 54 & 55 & 56 & 55 \\
 & Reasoning & 100 & 95 & 94 & 100 & 89 & 87 \\
 & Randomizer & 50 & 50 & 50 & 50 & 56 & 59 \\
\addlinespace
Claude Haiku 5.5 & Reasoning & 97 & 97 & 90 & 89 & 69 & 73 \\
 & Randomizer & 49 & 48 & 52 & 51 & 58 & 81 \\
\addlinespace
GPT-6 Luna & Reasoning & 97 & 96 & 95 & 68 & 57 & 66 \\
 & Randomizer & 52 & 53 & 52 & 50 & 56 & 74 \\
\addlinespace
Gemini 3.8 Flash & Reasoning & 94 & 57 & 98 & 50 & 100 & 100 \\
 & Randomizer & 50 & 51 & 66 & 51 & 93 & 99 \\
\addlinespace
Equilibrium &  & 50.0 & 50.0 & 50.0 & 50.0 & 62.5 & 78.1 \\
\bottomrule
\end{tabular*}
\begin{minipage}{\textwidth}
\vspace{0.4em}\footnotesize \textit{Notes.} Entries are the percentage of pairs of replies to the same prompt version, a variant at one base prize, that choose the same action, estimated without replacement within each version and averaged over versions. Each reply is an independent request, so each stands for a separate delegate. In equilibrium the rate is $p^2+(1-p)^2$ at the equilibrium probability $p$. Immediate answers and reasoning are at $T=1$, or the default temperature for Claude Haiku~5.5. The randomizer rows are the panel runs.
\end{minipage}
\end{table}

\clearpage
\section{Choice Histories in the Stated-Strategy Task}\label{app:history}

In the history treatment, each model makes 10 decisions in a row under the same prescribed probabilities and sees its earlier choices. The control runs the same sequences without the record. Against the same sequences without the record, the record moves the $\pi$-named action's share at $\pi=1/2$ from 74 to 50 percent for Gemma~4, from 66 to 50 percent for Qwen3.8, from 63 to 51 percent for Haiku, from 94 to 55 percent for Luna, and from 37 to 49 percent for Gemini. These five models get there by alternating, switching actions in 80 to 99 percent of consecutive decisions, so each choice is predictable from the one before. Nemotron switches in only 41 percent, and the record moves its share only from 88 to 65 percent. In 26 of its 48 sequences it chooses the $\pi$-named action 8 or 9 times in 10.

The alternating sequences are also too regular. At $\pi=1/2$ the share of the $\pi$-named action varies across sequences only a fifth to a third as much as under exact implementation for the frontier models, and hardly at all for the open-weight models. All 48 of Gemma~4's sequences and 46 of Qwen3.8's contain exactly five of each action.

At unequal probabilities, alternation cannot produce the target, and the record leaves the play of the frontier models and Nemotron nearly pure. Across 960 such decisions, Haiku chooses the less likely action 5 times, Luna 25 times, and Gemini 17 times, against 166 under exact implementation, and Nemotron chooses it 84 times in 3,839, against 666. Gemma~4 and Qwen3.8 instead use the record to fill a quota. Gemma~4 chooses the less likely action in 361 of 3,840 such decisions and Qwen3.8 in 348 of 3,838, against 666 under exact implementation, and almost never in the first decision of a sequence. At $\pi=1/3$, Gemma~4 chooses the $\pi$-named action exactly three times in 44 of its 48 ten-decision sequences and Qwen3.8 in 41, whereas exact implementation would give exactly three in only about 12, since its count varies from one sequence to the next.

\end{document}